\documentclass[%
 reprint,
 amsmath,amssymb,
 aps,
 prd
floatfix,
showkeys
]{revtex4-1}
\usepackage[utf8x]{inputenc} 
\usepackage{bm}
\usepackage{amssymb}
\usepackage{amsmath}
\usepackage{graphicx}
\usepackage{epsfig}
\usepackage{array}
\usepackage{float}
\usepackage{color}
\usepackage[usenames,dvipsnames]{xcolor}
\usepackage{epstopdf}
\usepackage{natbib}
\usepackage{lineno}
\usepackage{slashed}
\usepackage{color}
\usepackage{caption}
\usepackage{hyperref}
\usepackage{slashed}
\usepackage{soul}
\usepackage[bottom]{footmisc}

\newcommand{\sign}[1]{\text{sign}(#1)}
\newcommand{\para}{\parallel}
\newcommand{\Traza}[1]{\text{Tr}\left\{#1\right\}}

\newcommand{\be}{\begin{equation}}
\newcommand{\ee}{\end{equation}}
\newcommand{\bea}{\begin{eqnarray}}
\newcommand{\eea}{\end{eqnarray}}
\newcommand{\beas}{\begin{eqnarray*}}
\newcommand{\eeas}{\end{eqnarray*}}
\newcommand{\nn}{\nonumber}
\newcommand{\p}{\parallel}
\newcommand{\pp}{p_\parallel}

\newcommand{\ps}{\slashed{p}}

\newcommand{\Op}{\mathcal{O}}

\newcommand{\eB}{\left |  q_fB\right |}

\newcommand{\pt}{p_\perp}
\newcommand{\Bf}{\mathfrak{B}}

\begin{document}

\title{Gluon polarization tensor and dispersion relation in a weakly magnetized medium}


\author{Alejandro Ayala$^{1,2}$, Jorge David Casta\~no Yepes$^1$, L. A. Hern\'andez$^{1,2,3}$, Jordi Salinas$^1$, and R. Zamora$^{4,5}$ }
\affiliation{%
$^1$Instituto de Ciencias Nucleares, Universidad Nacional Aut\'onoma de M\'exico, Apartado Postal 70-543, CdMx 04510, Mexico.\\
$^2$Centre for Theoretical and Mathematical Physics, and Department of Physics, University of Cape Town, Rondebosch 7700, South Africa.\\
$^3$Facultad de Ciencias de la Educaci\'on, Universidad Aut\'onoma de Tlaxcala, Tlaxcala, 90000, Mexico.\\
$^4$Instituto de Ciencias B\'asicas, Universidad Diego Portales, Casilla 298-V, Santiago, Chile.\\
$^5$Centro de Investigaci\'on y Desarrollo en Ciencias Aeroespaciales (CIDCA), Fuerza A\'erea de Chile,  Santiago, Chile.}%


\begin{abstract}
We study the polarization and dispersion properties of gluons moving within a weakly magnetized background at one-loop order. To this end, we show two alternative derivations of the charged fermion propagator in the weak field expansion and use this expression to compute the lowest order magnetic field correction to the gluon polarization tensor. We explicitly show that, in spite of its cumbersome appearance, the gluon polarization tensor is transverse as required by gauge invariance. We also show that none of the three polarization modes develops a magnetic mass and that gluons propagate along the light cone, non withstanding that Lorentz invariance is lost due to the presence of the magnetic field. By comparing with the expression for the gluon polarization tensor valid to all orders in the magnetic field, the existence of a second solution, corresponding to a finite gluon mass, is shown to be spurious and an artifact of the lowest order approximation in the field strength. We also study the strength of the polarization modes for real gluons. We conclude that, provided the spurious solutions are discarded, the lowest order approximation to the gluon polarization and dispersion properties is good as long as the field strength is small compared to the loop fermion mass. 

\end{abstract}


\maketitle

\section{Introduction}\label{sec1}

In recent years, the properties of strongly interacting matter in the presence of magnetic fields have received a great deal of attention. Motivated by the possibility that magnetic fields of a large intensity --albeit short-lived-- can be produced in peripheral heavy-ion collisions at high energies, efforts, from the experimental as well as from the theoretical points of view, have been devoted to identify modifications induced on the propagation properties of quarks, gluons, and even hadrons in a magnetized medium. For instance, attempts have been made to link the observation of the charge separation along the magnetic field direction~\cite{STAR} with the chiral magnetic effect~\cite{CME} or the anomalous excess of the yield and $v_2$ of direct photons~\cite{experimentsyield,experimentsv2} with contributions from channels otherwise not present in the absence of a magnetic field~\cite{photons,Skokov,Zakharov,Tuchin}. More recently, measurements of a different global polarization of $\Lambda$ and $\overline{\Lambda}$, as the collision energy decreases~\cite{STARpol}, have been associated with the magnetic field produced in the reaction~\cite{magexpl}. Also, lattice QCD (LQCD) calculations~\cite{LQCD} have shown that for temperatures above the chiral restoration temperature, the quark-antiquark condensate decreases and that this temperature itself also decreases as the field intensity increases. This is the so-called 
{\it inverse magnetic catalysis} phenomenon and the search for its origin has been also intensively studied~\cite{Bruckmann,Farias,Ferreira,Ayala0, Ayala1,Ayala2,Ayala3,Avancini,Ayala4,vertex1,vertex2,Mueller}. Moreover, LQCD has also shown that the magnetic field-driven modifications of neutral and charged mesons are different~\cite{PhysRevD.97.034505}, with the neutral pion mass decreasing as the field intensity increases. In the linear sigma model, a possible explanation is found in the behavior of the coupling constants as function of the field strength~\cite{pionmassmag}.

When the magnetic field is the largest of the energy (squared) scales, a commonly used description to include its effects is the {\it strong field approximation}, whereby the propagation of charged particles occurs only within the Lowest Landau Level (LLL). The rationale behind this approximation is that fluctuations, whose typical energy is smaller than the separation between Landau Levels (proportional to the squared root of the field intensity), cannot induce transitions. This approach has been implemented in a recent calculation of the thermal corrections to the Debye magnetic mass~\cite{Debye}. However, there are many physical situations where the magnetic field may not be the largest of the energy scales. This is for example the case after the very early stages of a peripheral heavy-ion collision, since the field intensity is a fast decreasing function of time and thermal processes become important. In these situations another approximation, where the magnetic field is a small energy scale, is better suited. This scenario is best described by considering the propagation of charged particles in the {\it weak field approximation}. In this context, a recent calculation for the dissociation of heavy quarkonia in a weak magnetic field has been presented in Ref.~\cite{Hasan}.

The expressions for the charged fermion and scalar propagators in the weak field approximation were first obtained in Refs.~\cite{Chyi, Ayala-prop}, respectively. These propagators can be used, in particular, to find the quantum corrections to the propagation properties of neutral particles in a magnetized medium. In this work we concentrate on the computation of the gluon polarization tensor in the presence of a magnetic field at one-loop level using the weak field approximation to describe the fermion propagator. As we will show, the computation is plagued with subtleties, most notably, the enforcement of gauge invariance encoded in the transversality properties of the polarization tensor. We show that a systematic treatment of the weak field approximation ensures that spurious terms, that do not respect gauge invariance, vanish. The result is checked against the one obtained from the expansion to lowest non-trivial order in the field intensity of the general expression, which has been recently obtained in Ref.~\cite{polgen}, and we find a coincident result between both approaches. The method hereby presented is better suited than an expression valid to all orders in the magnetic field, to be extended at finite temperature and density, a relevant scenario, given the current and future explorations of the QCD phase diagram in heavy-ion experiments in the STAR-BES, FAIR, and NICA experiments. 

The work is organized as follows: In order to better identify the approximations that lead to the result for the charged fermion propagator in the weak field approximation obtained in Ref.~\cite{Chyi}, in Sec.~\ref{secII} we present two different weak-field expansion methods to derive such propagator. We show the explicit result of this expansion up to ${\mathcal{O}}(B^6)$. In Sec.~\ref{secIII} we obtain the expression for the gluon polarization tensor in the weak field approximation to second order in the field intensity. We compare with the expression obtained from the expansion to the same order obtained in Ref.~\cite{polgen} and verify the gauge invariance of the result. In Sec.~\ref{secIV} we analyse the dispersion relation for the three different propagating modes from the coefficients of their tensor structures. We show that, in spite of their cumbersome appearance, the gluon does not develop a magnetic mass in any mode and moves along the light cone. We also study the strength of the polarization modes for on-shell gluons. Finally, we summarize and conclude in Sec.~\ref{concl} leaving for the appendix the explicit derivation of the charged fermion propagator in the weak field expansion using the second method mentioned above.

\section{Fermion propagator in the weak field limit} \label{secII}

In order to compute the gluon polarization tensor in the presence of a weak magnetic field, we first proceed to obtain the expression for the charged fermion propagator that appears in the one-loop correction to the gluon polarization tensor in this limit.
This object has been first obtained in Ref.~\cite{Chyi}, Nevertheless, we hereby present two alternative methods to obtain it. For the first method, we make a direct expansion in powers of the field strength starting from the full fermion propagator using Schwinger's proper-time representation. For the second method, we start from the full expression of the fermion propagator written in terms of an expansion of Landau Levels. In order to set the stage for the calculation, we first describe the notation and conventions to be used throughout the calculations.

\subsection{Notation}

Before we show the derivation of the fermion propagator in the weak field limit, here we spell out the notation and conventions that we use in this work. The magnetic field is taken as pointing  along the $\hat{\mathbf{z}}$-axis and having an intensity $B=|{\mathbf{B}}|$, namely, ${\mathbf{B}}=B\hat{\mathbf{z}}$. The field couples to fermions of mass $m_f$ through their electric charge $q_f$. In addition, we use the following notation and conventions:
\begin{itemize}
    \item {\slshape Metric tensor}
    \begin{gather}
        g_{\mu \nu}=g^{\mu \nu}=\begin{pmatrix} 1 & 0 & 0 & 0 \\ 0 & -1 & 0 & 0 \\ 0 & 0 & -1 & 0 \\ 0 & 0 & 0 & -1 \end{pmatrix}.
        \label{metrictensor}
    \end{gather}
     \item {\slshape Parallel and transverse metric tensors}
    \begin{equation}
        g^{\mu \nu}=g^{\mu \nu}_\parallel+g^{\mu \nu}_\perp,
        \label{metricparaperp}
    \end{equation}
    with
    \begin{gather}
        g^{\mu \nu}_\parallel=\begin{pmatrix} 1 & 0 & 0 & 0 \\ 0 & 0 & 0 & 0 \\ 0 & 0 & 0 & 0 \\ 0 & 0 & 0 & -1 \end{pmatrix},
        \label{metrictensorpara}
    \end{gather}
    and
    \begin{gather}
        g^{\mu \nu}_\perp=\begin{pmatrix} 0 & 0 & 0 & 0 \\ 0 & -1 & 0 & 0 \\ 0 & 0 & -1 & 0 \\ 0 & 0 & 0 & 0 \end{pmatrix}.
        \label{metrictensorperp}
    \end{gather}
    \item {\slshape Four-vectors}
    \begin{align}
        X^\mu&=(X^\mu_\parallel+X^\mu_\perp), \nonumber \\
        X_\mu&=(X_\mu^\parallel-X_\mu^\perp),
        \label{fourvector}
    \end{align}
    with
    \begin{align}
        X^\mu_\parallel&=(X_0,0,0,X_3), \ \ \ \ X_\mu^\parallel=(X_0,0,0,-X_3), \nonumber \\
        X^\mu_\perp&=(0,X_1,X_2,0), \ \ \ \ X_\mu^\perp=(0,X_1,X_2,0).
        \label{fourvectorcomponents}
    \end{align}
    \item{\slshape Squared four-vector}
    \begin{align}
        X^\mu X_\mu&=X^2=X_\parallel^2-X_\perp^2 \nonumber \\
        &=(X_0^2-X_3^2)-(X_1^2+X_2^2).
        \label{squaredvector}
    \end{align}
\end{itemize}

\subsection{Weak field approximation from Schwinger formalism}

The most general expression for the charged fermion propagator in the presence of a constant and uniform magnetic field is
\begin{equation}
    iS(x,x')=e^{i\Phi (x,x')}\int \frac{d^4 p}{(2\pi)^4} e^{i (x-x')p} \ iS(p),
    \label{fullprop}
\end{equation}
where $\Phi(x,x')$ is the so-called Schwinger's phase factor and $iS(p)$ is the translationally invariant term. Since, we are interested in computing the one-loop correction to the gluon polarization tensor, the phase factor vanishes and thus we hereby do not account for it. 

We express $iS(p)$ using the Schwinger proper-time representation
\begin{align}
    iS(p)&=i\int_0^\infty d\tau e^{\tau (p_\parallel^2-p_\perp^2 \frac{\tanh(\tau |q_fB|)}{\tau |q_fB|}-m_f^2)} \nonumber \\
    &\times \Big\{ [m_f+\slashed{p}_\parallel][1+i\gamma^1 \gamma^2 \tanh(\tau |q_fB|)\text{sign}(q_fB)]  \nonumber \\
    &-\frac{\slashed{p}_\perp}{\cosh(\tau |q_fB|)^2}\Big \}.
    \label{Schwingerprop}
\end{align}

Since in this work $|q_fB|$ is the smallest energy (squared) scale, in order to find a suitable representation of the propagator in the weak field approximation, we now proceed to perform a Taylor's series expansion of Eq.~(\ref{Schwingerprop}) for $\tau |q_fB|\sim 0$ and then to integrate over the proper time parameter $\tau$. It can be shown by using this straightforward expansion that the propagator, written here up to ${\mathcal{O}}(B^6)$, becomes
\begin{eqnarray}
    iS(p)&\simeq& i\frac{m_f+\slashed{p}}{p^2-m_f^2}- |q_fB|\gamma^1 \gamma^2\frac{m_f+\slashed{p}_\parallel}{(p^2-m_f^2)^2}\text{sign}(q_fB)\nonumber \\
    &-&2i|q_fB|^2\frac{(m_f^2-p_\parallel^2)\slashed{p}_\perp+p_\perp^2(m_f+\slashed{p}_\parallel)}{(p^2-m_f^2)^4}\nonumber \\
    &+&2|q_fB|^3\gamma^1 \gamma^2\frac{(m_f+\slashed{p}_\parallel)(p_\parallel^2+3p_\perp^2-m_f^2)}{(p^2-m_f^2)^5}\text{sign}(q_fB)\nonumber \\
    &-&8i|q_fB|^4\frac{(2p_\parallel^2+3p_\perp^2-2m_f^2)}{(p^2-m_f^2)^7}\nonumber \\
    &\times& ((m_f^2-p_\parallel^2)\slashed{p}_\perp+p_\perp^2(m_f+\slashed{p}_\parallel)) \nonumber \\
    &-&8|q_fB|^5\gamma^1\gamma^2\frac{(m_f+\slashed{p}_\parallel)}{(p^2-m_f^2)^8}\text{sign}(q_fB)\nonumber \\
    &\times& (18p_\perp^2(p_\parallel^2-m_f^2)+2(m_f^2-p_\parallel^2)^2+15p_\perp^4)\nonumber \\
    &+&16i|q_fB|^6\frac{((m_f^2-p_\parallel^2)\slashed{p}_\perp+p_\perp^2(m_f+\slashed{p}_\parallel))}{(p^2-m_f^2)^{10}}\nonumber \\
    &\times& (78p_\perp^2(p_\parallel^2-m_f^2)+17(m_f^2-p_\parallel^2)^2+45p_\perp^4).
    \label{fromproptime}
\end{eqnarray}
We notice that Eq.~(\ref{fromproptime}) coincides with the weak field expansion performed in Ref.~\cite{Chyi}. It may be surprising that a simple Taylor's series expansion, starting out from Schwinger's proper time representation, yields the same result than the one in Ref.~\cite{Chyi}, where the result was obtained using a much more complicated method. The computation also reveals that the dominant region for the integration over the proper-time parameter is the small $\tau$ region and that, in order to obtain analytical results, this region can be extended to cover the whole original domain $0\leq\tau \leq\infty$.

\subsection{Weak field approximation from the Landau Levels representation}
To find an expression of the fermion propagator in a weak magnetic field, we can also start from the expression of the full propagator  
written as a sum over Landau Levels~\cite{Miransky1,Miransky2}
\bea
\!\!\!\!\!\! iS(p)=ie^{-\pt^2/\eB}\sum_{n=0}^{\infty}(-1)^n\frac{\mathcal{D}_n(q_fB,p)}{p_\p^2-m_f^2-2n\left |  q_fB\right |}.
\label{eq:MiranskyPropagator}
\eea
The factor $\mathcal{D}_n(q_fB,p)$ is defined as
\bea
\mathcal{D}_n(q_fB,p)&=&2(\ps_\p+m_f)\mathcal{O}^{-}L_n^0\left(\frac{2\pt^2}{\left | q_fB \right |}\right)\nn\\
&-&2(\ps_\p+m_f)\mathcal{O}^{+}L_{n-1}^0\left(\frac{2\pt^2}{\left | q_fB \right |}\right)\nn\\
&+&4\ps_\perp L_{n-1}^1\left(\frac{2\pt^2}{\left | q_fB \right |}\right),
\label{Dndef}
\eea
where $L_n^{m}(x)$ are the generalized Laguerre polynomials and
\bea
\mathcal{O}^{\pm}\equiv\frac{1}{2}\left[1\pm i\ \sign{q_fB}\gamma^1\gamma^2\right].
\eea

The desired approximation has to take into account the contribution from all Landau Levels. If we assume that the magnetic field is the weakest energy scale, a formal series representation for the fermion propagator in powers of $|q_fB|$ can be computed. For this purpose, notice that the denominators of Eq.~(\ref{eq:MiranskyPropagator}) admit a geometrical series expansion, namely,
\bea
\frac{1}{\pp^2-m_f^2-2n\eB}=\frac{1}{\pp^2-m_f^2}\sum_{k=0}^\infty\left(\frac{2n\eB}{\pp^2-m_f^2}\right)^k.\nn\\
\label{DenominatortoGeomSum}
\eea

From the above, the summation over the Landau level index $n$ becomes
\bea
\mathcal{S}_k=\sum_{n=0}^{\infty}(-1)^n n^k L_n^m(2\alpha),
\eea
where $m=0$ or $1$ and $\alpha=\pt^2/eB$. To carry out this kind of sum, we use the identities
\begin{subequations}
\bea
e^{-\alpha}\sum_{n=0}^\infty(-1)^n e^{-2inv}L_n^0(2\alpha)=\frac{e^{iv}}{2\cos v}e^{-i\alpha\tan v},\nn\\
\eea
\bea
e^{-\alpha}\sum_{n=0}^{\infty}(-1)^n e^{-2i(n+1)v}L_{n}^1(2\alpha)=\frac{1}{4\cos^2v}e^{-i\alpha\tan v},\nn\\
\eea
\label{Identities1def}
\end{subequations}
so that
\begin{subequations}
\bea
&&e^{-\alpha}\sum_{n=0}^\infty(-1)^n n^k L_n^0(2\alpha)\nn\\
&=&\frac{1}{(-2i)^k}\lim_{v\rightarrow0}\frac{\partial^k}{\partial v^k}\left(\frac{e^{iv}}{2\cos v}e^{-i\alpha\tan v}\right),
\eea
and
\bea
&&e^{-\alpha}\sum_{n=0}^{\infty}(-1)^n (n+1)^ke^{-2i(n+1)v}L_{n}^1(2\alpha)\nn\\
&=&\frac{1}{(-2i)^k}\lim_{v\rightarrow0}\frac{\partial^k}{\partial v^k}\left(\frac{1}{4\cos^2v}e^{-i\alpha\tan v}\right).
\eea
\label{DerivativeIdentities1}
\end{subequations}

Equations~(\ref{DerivativeIdentities1}) provide the expression for the fermion propagator up to any given order in $|q_fB|$. For instance, up to ${\mathcal{O}}(B^6)$, we obtain
\bea
iS(p)&\simeq& i\frac{m_f+\slashed{p}}{p^2-m_f^2}- |q_fB|\gamma^1 \gamma^2\frac{m_f+\slashed{p}_\parallel}{(p^2-m_f^2)^2}\text{sign}(q_fB)\nonumber \\
&-&2i|q_fB|^2\frac{(m_f^2-p_\parallel^2)\slashed{p}_\perp+p_\perp^2(m_f+\slashed{p}_\parallel)}{(p^2-m_f^2)^4}\nonumber \\
&+&2|q_fB|^3\gamma^1 \gamma^2\frac{(m_f+\slashed{p}_\parallel)(p_\parallel^2+3p_\perp^2-m_f^2)}{(p^2-m_f^2)^5}\text{sign}(q_fB)\nonumber \\
&-&8i|q_fB|^4\frac{(2p_\parallel^2+3p_\perp^2-2m_f^2)}{(p^2-m_f^2)^7}\nonumber \\
&\times& ((m_f^2-p_\parallel^2)\slashed{p}_\perp+p_\perp^2(m_f+\slashed{p}_\parallel)) \nonumber \\
&-&8|q_fB|^5\gamma^1\gamma^2\frac{(m_f+\slashed{p}_\parallel)}{(p^2-m_f^2)^8}\text{sign}(q_fB)\nonumber \\
&\times& (18p_\perp^2(p_\parallel^2-m_f^2)+2(m_f^2-p_\parallel^2)^2+15p_\perp^4) \nonumber \\
&+&16i|q_fB|^6\frac{((m_f^2-p_\parallel^2)\slashed{p}_\perp+p_\perp^2(m_f+\slashed{p}_\parallel))}{(p^2-m_f^2)^{10}}\nonumber \\
&\times& (78p_\perp^2(p_\parallel^2-m_f^2)+17(m_f^2-p_\parallel^2)^2+45p_\perp^4).
\label{iSfinalfromLandau}
\eea
For details see Appendix~\ref{ApDerivationsFromLandau}. Equation~(\ref{iSfinalfromLandau}) coincides with Eq.~(\ref{fromproptime}). The computation also reveals that the approximation is valid provided $|q_fB|<m_f^2$, as can be seen from Eq.~(\ref{DenominatortoGeomSum}) where, in order for the geometric series to converge, independent of the value of $p_\parallel^2$, this condition needs to be satisfied. The above is clearer when $p_\parallel^2$ is continued to Eucledian values $p_\parallel^2\to -p^E_\parallel$, which is a step previous to using this propagator in the context of finite temperature calculations using the Matsubara formalism. 

\section{Gluon polarization tensor}\label{secIII}

\begin{figure}[b!]
    \centering
    \includegraphics[scale=0.3]{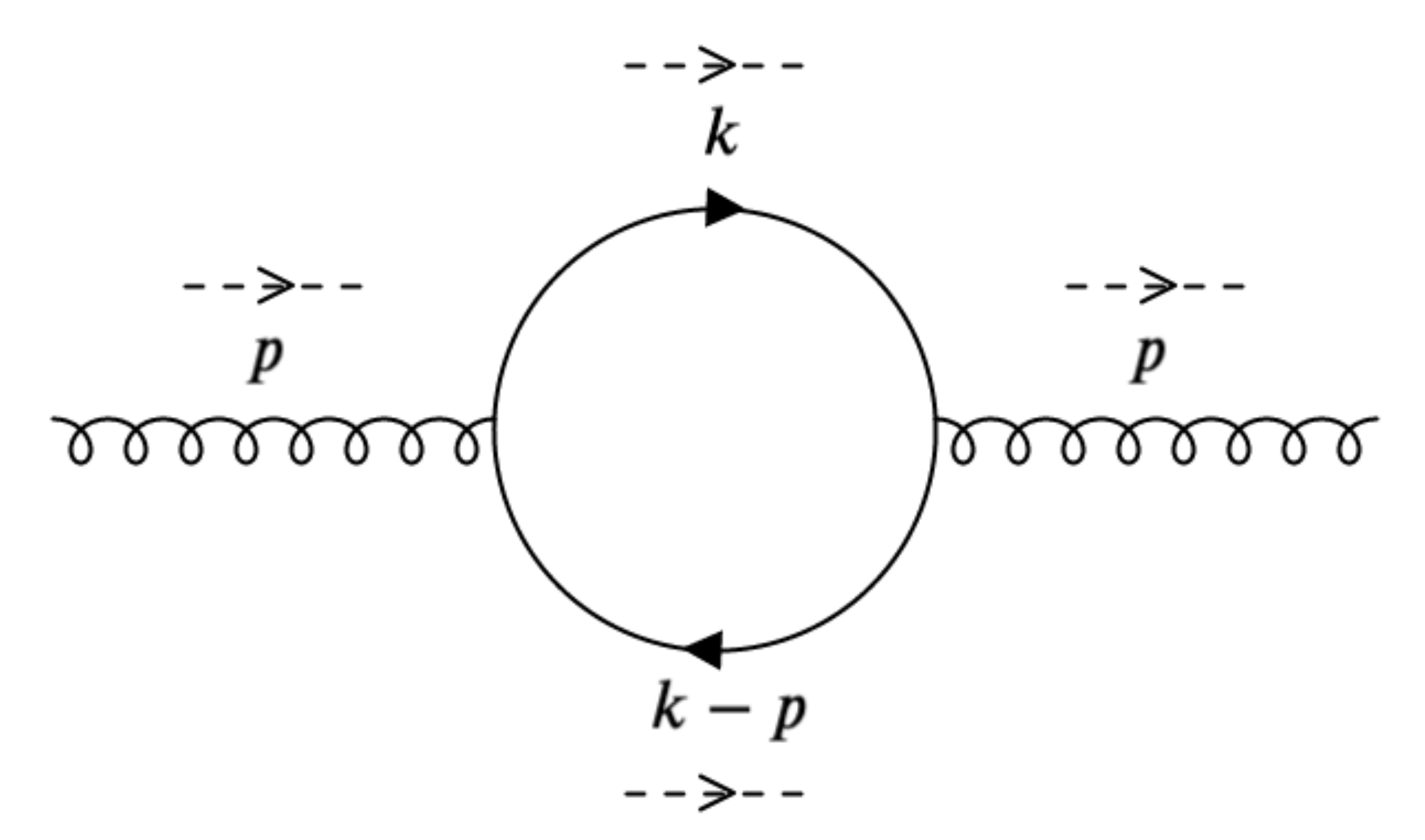}
    \caption{One-loop diagram representing the gluon polarization tensor.}
    \label{fig1}
\end{figure}

With the expression for a charged fermion propagator in powers of the field strength at hand, we are in the position to compute the gluon polarization tensor. At one-loop order, this tensor is obtained from the Feynman diagram depicted in Fig.~\ref{fig1}, where the loop is made out of a quark-antiquark pair. Its explicit expression is
\begin{align}
    i\Pi^{\mu \nu}_{ab}(p)=&-\int \frac{d^4 k}{(2\pi)^4} \text{Tr} \{ igt_b\gamma^\nu iS(k)igt_a\gamma^\mu iS(k-p) \} \nonumber \\
    &+ \text{C.C.},
\end{align}
where $igt_a\gamma^\mu$ is the QCD quark-gluon vertex. The global negative sign comes from the fermion loop and C.C. refers to the charge conjugate contribution. Here, we restrict ourselves to using the expansion for $iS$ only up to  ${\mathcal{O}(B^2)}$, using either Eq.~(\ref{iSfinalfromLandau}) or Eq.~(\ref{fromproptime}), namely,
\begin{align}
iS(k)&=i \frac{\slashed{k}+m_f}{k^2-m^2}-|q_fB|\frac{\gamma^1 \gamma^2(\slashed{k}_\parallel+m_f)}{(k^2-m_f^2)^2}\text{sign}(q_fB)\nonumber \\ &-|q_fB|^2\frac{2ik_\perp^2}{(k^2-m_f^2)^4}\Bigg[ m_f+\slashed{k}_\parallel+\slashed{k}_\perp \Bigg(\frac{m_f^2-k_\parallel^2}{k_\perp^2}\Bigg) \Bigg] \nonumber \\ \nonumber \\
&\equiv iS^{(0)}(k)+iS^{(1)}(k)+iS^{(2)}(k).
\label{propeB2}
\end{align}

Gauge invariance requires that the gluon polarization tensor be transverse. However, the breaking of Lorentz symmetry makes this tensor to split into three transverse structures, such that the gluon polarization tensor (omitting the diagonal color factor $\delta_{ab}$) can be written as~\cite{Hattori1} (see also Ref.~\cite{Aritra})
\begin{equation}
 \Pi^{\mu \nu}= P_\parallel \mathcal{P}_{\parallel}^{\mu \nu}+P_\perp \mathcal{P}_{\perp}^{\mu \nu}+P_0 \mathcal{P}_{0}^{\mu\nu},
 \label{basistensor}
\end{equation}
where
\begin{align}
  \mathcal{P}_{\parallel}^{\mu\nu}  &= g^{\mu\nu}_\parallel-\frac{q^\mu_\parallel q^\nu_\parallel}{q^2_\parallel}, \nonumber \\
  \mathcal{P}_{\perp}^{\mu\nu}  &= g^{\mu\nu}_\perp+\frac{q^\mu_\perp q^\nu_\perp}{q^2_\perp}, \nonumber \\
  \mathcal{P}_{0}^{\mu\nu}&=g^{\mu\nu}-\frac{q^\mu q^\nu}{q^2}-(\mathcal{P}_{\parallel}^{\mu\nu}+\mathcal{P}_{\perp}^{\mu\nu}).
   \label{elembasis}
\end{align}
Nevertheless, the explicit computation reveals that the polarization tensor exhibits in addition a dependence proportional to the non-transverse tensor structures
$g^{\mu \nu}_\parallel$ and $g^{\mu \nu}_\perp$, 
whose coefficients do not obviously vanish. Thus, the tensor needs to be written in general as
\begin{equation}
    \Pi^{\mu \nu}= P_\parallel \mathcal{P}_{\parallel}^{\mu \nu}+P_\perp \mathcal{P}_{\perp}^{\mu \nu}+P_0 \mathcal{P}_{0}^{\mu\nu}+A_1 g^{\mu \nu}_\parallel +A_2 g^{\mu \nu}_\perp. \label{eq24}
\end{equation}

In this work we show that the coefficients $A_1$ and $A_2$ do vanish, and thus that the gluon polarization tensor obtained using the fermion propagator expanded in powers of $|q_fB|$ up to second order is indeed a transverse tensor. This result was also obtained in Ref.~\cite{polgen} from an expansion to the same order in $|q_fB|$ of the general expression for the polarization tensor in the presence of a magnetic field of arbitrary intensity.

\begin{figure}[b]
    \centering
    \includegraphics[scale=0.5]{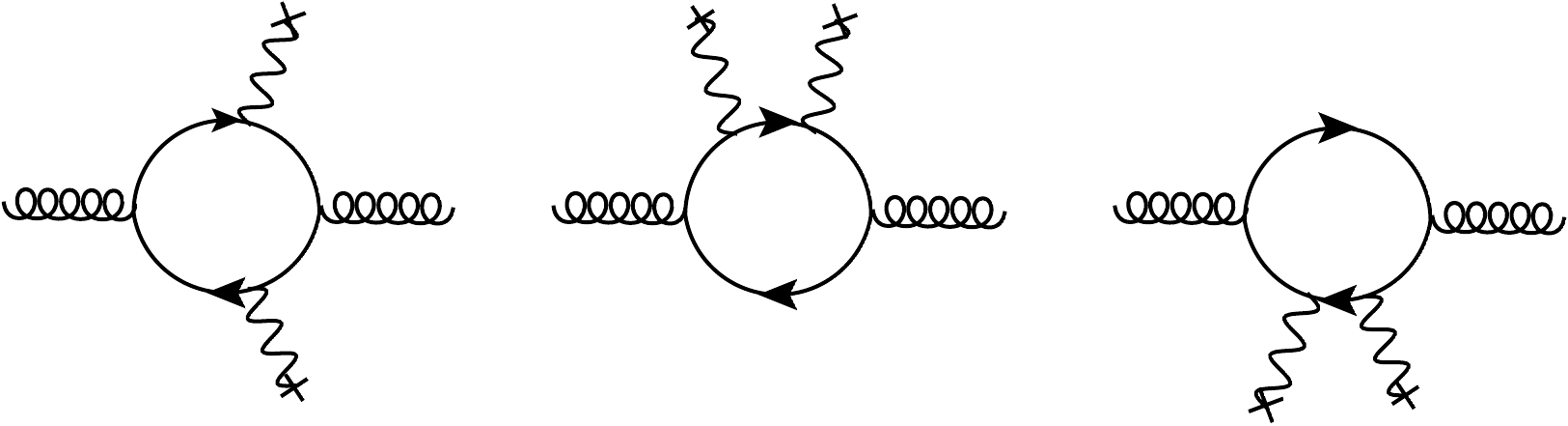}
    \caption{Feynman diagrams contributing to the one-loop gluon polarization tensor in the weak field limit to second order in $B$.}
    \label{fig2}
\end{figure}

When computing the gluon polarization tensor at one-loop order, the internal quark/antiquark lines are described by a fermion propagator given by Eq.~(\ref{propeB2}). Since a diagram with a single external field insertion vanishes, as required by Furry's theorem, the first nontrivial magnetic contribution to the gluon polarization tensor turns out to be of order $\mathcal{O}(B^2)$. The relevant Feynman diagrams to be computed are depicted in Fig.~\ref{fig2}. The photon lines represent the coupling of the external magnetic field to the quark-antiquark pair in the loop. There are two kinds of contributions at order $\mathcal{O}(B
^2)$: The first one comes from the product of the linear terms in Eq.~(\ref{propeB2}), and the second one results from the product of the vacuum and the quadratic terms. We call the first term $\Pi^{\mu \nu}_{(1,1)}$ and the second term $\Pi^{\mu\nu}_{(2,0)}+\Pi^{\mu\nu}_{(0,2)}$. $\Pi^{\mu \nu}_{(1,1)}$ is its own \text{C.C.} and thus there is no need that its \text{C.C.} is added up. On the other hand, $\Pi^{\mu\nu}_{(2,0)}$ and $\Pi^{\mu\nu}_{(0,2)}$ are the \text{C.C.} of each other and thus both contributions need to be taken into account. Given the subtleties involved in the calculation, we now proceed to compute in detail each one of these terms. 

\subsection{\texorpdfstring{$\Pi^{\mu \nu}_{(1,1)}$}{} contribution}

The contribution from $\Pi^{\mu \nu}_{(1,1)}$ can written as
\begin{align}
    i\Pi^{\mu \nu}_{(1,1)}&=-\int \frac{d^4k}{(2 \pi)^4} \nonumber \\
    &\times \frac{\Traza {igt_b \gamma^\nu iS^{(1)}(k-p)igt_a \gamma^\mu iS^{(1)}(k)}}{(k^2-m_f^2)^2\left[(k-p)^2-m_f^2\right]^2} \nonumber \\
   &=2\eB^2 g^2 \int \frac{d^4k}{(2 \pi)^4} \nonumber \\
   &\times  \frac{1}{(k^2-m_f^2)^2\left[(k-p)^2-m_f^2\right]^2} \nonumber \\
   &\times  \Biggl[ k_\parallel^\nu(k_\parallel^\mu-p_\parallel^\mu)+k_\parallel^\mu(k_\parallel^\nu-p_\parallel^\nu) \nonumber \\
    &-(g_\parallel^{\mu\nu}-g_\perp^{\mu\nu})\left(k_\parallel \cdot\left(k_\parallel-p_\parallel\right) -m_f^2 \right)  \Biggr].
    \label{pi11initial}
\end{align}

To carry out the calculation, we introduce two Schwinger parameters $x_1$ and $x_2$, using the general expression
\begin{equation}
   \frac{1}{k^2-m^2}=-\int_0^\infty dx \ e^{x(k^2-m^2)}.
\label{idenSchwinger}
\end{equation}
Thus, using Eq.~(\ref{idenSchwinger}), we rewrite the denominators in Eq.~(\ref{pi11initial}) as 
\begin{eqnarray}
\!\!\!\!\!\!\!\!&& \frac{1}{(k^2-m_f^2)^2\left[(k-p)^2-m_f^2\right]^2}\nonumber \\
&=& \frac{1}{(k^2-m_{1}^2)^2\left[(k-p)^2-m_{2}^2\right]^2} \nonumber \\
&=&\frac{\partial}{\partial m_{1}^2} \frac{\partial}{\partial m_{2}^2} \int d^2x \ e^{x_2(k^2-m_{1}^2)} e^{x_1((k-p)^2-m_{2}^2)},
\label{denominators}
\end{eqnarray}
where in order to exponentiate the denominators in a symmetric manner, we have distinguished the mass in each of the original factors to take the derivatives. We later bring the masses back to be the same.
Equation~(\ref{pi11initial}) can therefore be written as
\begin{eqnarray}
i\Pi^{\mu \nu}_{(1,1)} &=& 2\eB^2 g^2 \frac{\partial}{\partial m_{1}^2} \frac{\partial}{\partial m_{2}^2} \int \frac{d^4k}{(2 \pi)^4}  \nonumber \\
&\times& \int d^2x\,\exp\Bigl[(x_1+x_2)\left(k-\frac{x_1 }{x_1+x_2} p\right)^2 \nonumber \\
&+& \frac{x_1 x_2}{x_1+x_2}p^2-(m_{1}^2 x_1 + m_{2}^2 x_2) \Bigr] \nonumber \\
&\times& \Bigl\{ k_\parallel^\nu(k_\parallel^\mu-p_\parallel^\mu)+k_\parallel^\mu(k_\parallel^\nu-p_\parallel^\nu)
-(g_\parallel^{\mu\nu}-g_\perp^{\mu\nu}) \nonumber \\
&&\quad\times \left[ k_\parallel \cdot(k_\parallel-p_\parallel) -m_f^2 \right]\Bigr\}.
\end{eqnarray}
In order to perform the integral over the internal momenta, we make the change of variable
\begin{equation}
    l = \left(k-\frac{x_1 }{x_1+x_2} p\right), 
    \label{cambio}
\end{equation}
and by replacing $m_1=m_2=m_f$ we obtain
\begin{eqnarray}
i\Pi^{\mu \nu}_{(1,1)}&=& 2\eB^2 g^2 \int\frac{d^4l}{(2 \pi)^4}\int d^2x\, x_1 x_2\nonumber \\
&\times&\exp\Bigl[(x_1+x_2) (l^2-m_f^2) + \frac{x_1 x_2}{x_1+x_2}p^2 \Bigr] \nonumber \\
&\times& \Bigg\{ 2 l_\parallel^\mu l_\parallel^\nu - 2\frac{x_1 x_2}{(x_1+x_2)^2} p_\parallel^\mu p_\parallel^\nu 
-(g_\parallel^{\mu\nu}-g_\perp^{\mu\nu})
\nonumber \\
&&\quad\times\left[l_\parallel^2 -\frac{x_1 x_2}{(x_1+x_2)^2}p_\parallel^2 -m_f^2\right]\Bigg\}.
\end{eqnarray}
Using the tensor basis in Eq.~(\ref{elembasis}), we can write
\begin{equation}
p_\parallel^\mu p_\parallel^\nu = - p_\parallel^2 \mathcal{P}_\parallel^{\mu \nu} + p_\parallel^2 g_\parallel^{\mu \nu},
\end{equation}
to then obtain
\begin{eqnarray}
i\Pi^{\mu \nu}_{(1,1)} &=& 2\eB^2 g^2\int \frac{d^4l}{(2 \pi)^4}\int d^2x\, x_1 x_2  \nonumber \\
&\times&\exp\Bigl[(x_1+x_2) (l^2-m_f^2) + \frac{x_1 x_2}{x_1+x_2}p^2 \Bigr] \nonumber \\
&\times&\Bigg[- 2 l_\parallel^\mu l_\parallel^\nu - 2\frac{x_1 x_2}{(x_1+x_2)^2} p_\parallel^2  \mathcal{P}_\parallel^{\mu \nu} \nonumber \\
&&\;\; +g_\parallel^{\mu\nu}\left(m_f^2+l_\parallel^2 +\frac{x_1 x_2}{(x_1+x_2)^2}p_\parallel^2\right) \nonumber \\
&&\;\;-g_\perp^{\mu\nu}\left(m_f^2+l_\parallel^2 -\frac{x_1 x_2}{(x_1+x_2)^2}p_\parallel^2\right) \Bigg].
\end{eqnarray}
Also, to carry out the integration over the parallel components of the internal momentum, we use the substitution
\begin{equation}
l_\parallel^\mu  l_\parallel^\nu\to \frac{1}{2} g^{\mu \nu}_\parallel l_\parallel^2,
\end{equation}
after which the integration yields
\begin{eqnarray}
i\Pi^{\mu \nu}_{(1,1)} &=& -\frac{2i\eB^2 g^2}{16 \pi^2}  \int d^2x \frac{x_1 x_2}{(x_1+x_2)^4} \nonumber \\
&\times& \exp\Bigl[-(x_1+x_2)m_f^2 + \frac{x_1 x_2}{x_1+x_2}p^2 \Bigr] \nonumber \\
&\times& \Bigl\{ 
+g_\parallel^{\mu\nu}\left[m_f^2 (x_1+x_2)^2- x_1 x_2 p_\parallel^2\right] \nonumber \\
&&\;\;+g_\perp^{\mu\nu}\left[-m_f^2 (x_1+x_2)^2- (x_1+x_2) - x_1 x_2 p_\parallel^2\right]  \nonumber \\
&&\;\;+2 x_1 x_2  p_\parallel^2  \mathcal{P}_\parallel^{\mu \nu}\Bigr\}.
\end{eqnarray}
Finally, we perform the change of variables $x_1=s (1-y)$ and $x_2= s y$ to get
\begin{eqnarray}
i\Pi^{\mu \nu}_{(1,1)} &=& -\frac{2i\eB^2 g^2}{16 \pi^2}  \int_0^1 dy \int_0^\infty  ds \nonumber \\
&\times& s y (1-y) \exp\Bigl[-s(m_f^2 + y(y-1)p^2) \Bigr] \nonumber \\
&\times&\Bigg\{+2 y(1-y)  p_\parallel^2  \mathcal{P}_\parallel^{\mu \nu} \nonumber \\
&&\quad +g_\parallel^{\mu\nu}\left[m_f^2 -  y(1-y) p_\parallel^2\right] \nonumber \\
&&\quad+g_\perp^{\mu\nu}\left[-m_f^2 - \frac{1}{s} - y(1-y) p_\parallel^2\right] \Bigg\}.
\label{pi11withyands}
\end{eqnarray}
Equation~(\ref{pi11withyands}) represents the first contribution to the gluon polarization tensor. We now proceed to compute the remaining pieces.
\subsection{\texorpdfstring{$\Pi^{\mu \nu}_{(2,0)}$}{} and \texorpdfstring{$\Pi^{\mu \nu}_{(0,2)}$}{} contributions}

The remaining contribution to the polarization tensor of order $\mathcal{O}(B^2)$ can be split in two terms, $\Pi^{\mu \nu}_{(2,0)}$ and $\Pi^{\mu \nu}_{(0,2)}$. The first term corresponds to the second Feynman diagram depicted in Fig.~\ref{fig2}, and it has the following expression 
\bea
i\Pi_{(2,0)}^{\mu\nu} &=& -\int \frac{d^4k}{(2\pi)^4} \nonumber \\
&\times&\Traza{ igt_b\gamma^\nu iS^{(0)}(k-p)igt_a\gamma^\mu iS^{(2)}(k)} \nonumber \\
&=& -g^2\eB^2\int \frac{d^4k}{(2\pi)^4} \nonumber \\
&\times&\frac{1}{((k-p)^2-m_f^2)(k^2-m_f^2)^4} \nonumber \\
&&\times \text{Tr}\Big\{
\gamma^\nu\gamma^\alpha\gamma^\mu\gamma^\beta (k-p)_\alpha
\big[k_\beta^\perp(k^2-m_f^2)\nonumber \\
&&\quad -(k_\beta^\para-k_\beta^\perp)k_\perp^2\big]-\gamma^\nu\gamma^\mu k_\perp^2m_f^2\Big\}.
\eea
We proceed to compute this contribution in the same manner as for $\Pi^{\mu \nu}_{(1,1)}$. We use the relations in Eqs.~(\ref{idenSchwinger}) and~(\ref{denominators}) to introduce Schwinger parameters and rewrite the denominators. After the change of variables of Eq.~(\ref{cambio}) we obtain
\bea
i\Pi_{(2,0)}^{\mu\nu} &=&\eB^2 g^2 \Biggl[ \frac{1}{2!} \int \frac{d^4l}{(2 \pi)^4} \int d^2x\, x_2^2 \nonumber \\
 &&\times \exp\Bigl[(x_1+x_2) (l^2-m_f^2) + \frac{x_1 x_2}{x_1+x_2}p^2 \Bigr] \nonumber \\
 &\times& \text{Tr}\Big\{
\gamma^\nu\gamma^\alpha\gamma^\mu\gamma^\beta \left(l-\frac{x_2 p}{x_1+x_2}\right)_\alpha^\perp \nonumber \\
&\times&\left(l+\frac{x_1 p}{x_1+x_2}\right)_\beta^\perp\Big\} \nonumber \\
&-&\frac{1}{3!} \int \frac{d^4l}{(2 \pi)^4} \int_0^\infty dx_1 dx_2 \ x_2^3 \nonumber \\
 &&\times \exp\Bigl[(x_1+x_2) (l^2-m_f^2) + \frac{x_1 x_2}{x_1+x_2}p^2 \Bigr] \nonumber \\
&&+ \text{Tr} \Big\{
\gamma^\nu\gamma^\alpha\gamma^\mu\gamma^\beta \left(l-\frac{x_2 p}{x_1+x_2}\right)_\alpha \nonumber \\
&\times&\left(l+\frac{x_1 p}{x_1+x_2}\right)_\beta \left(l+\frac{x_1 p}{x_1+x_2}\right)_\perp^2\nonumber\\
&+&\gamma^\nu\gamma^\mu \left(l+\frac{x_1 p}{x_1+x_2}\right)_\perp^2m_f^2 \Big\} \Biggr]. 
\label{pi20l}
\eea

In order to compute the integral over the internal momenta in Eq.~(\ref{pi20l}), we split $\Pi_{(2,0)}^{\mu\nu}$ into three terms
\bea
i\Pi_{(2,0)}^{\mu\nu} &=&I_1^{\mu\nu}+I_2^{\mu\nu}+I_3^{\mu\nu},
\eea
where
\bea
 I_1^{\mu\nu} &=&\eB^2 g^2  \frac{1}{2!} \int \frac{d^4l}{(2 \pi)^4} \int_0^\infty dx_1 dx_2 \ x_2^2 \nonumber \\
 &&\times \exp\Bigl[(x_1+x_2) (l^2-m_f^2) + \frac{x_1 x_2}{x_1+x_2}p^2 \Bigr] \nonumber \\
 &\times& \text{Tr}\Big\{
\gamma^\nu\gamma^\alpha\gamma^\mu\gamma^\beta \left(l-\frac{x_2 p}{x_1+x_2}\right)_\alpha^\perp \left(l+\frac{x_1 p}{x_1+x_2}\right)_\beta^\perp\Big\}, \nonumber
\eea
\bea
I_2^{\mu\nu} &=&-\eB^2 g^2 \frac{1}{3!} \int \frac{d^4l}{(2 \pi)^4} \int_0^\infty dx_1 dx_2 \ x_2^3 \nonumber \\
 &&\times \exp\Bigl[(x_1+x_2) (l^2-m_f^2) + \frac{x_1 x_2}{x_1+x_2}p^2 \Bigr] \nonumber \\
&&+ \text{Tr}\Big\{
\gamma^\nu\gamma^\alpha\gamma^\mu\gamma^\beta \left(l-\frac{x_2 p}{x_1+x_2}\right)_\alpha \left(l+\frac{x_1 p}{x_1+x_2}\right)_\beta \nonumber \\ &&\times \left(l+\frac{x_1 p}{x_1+x_2}\right)_\perp^2\Big\}, \nonumber
\eea
\bea
I_3^{\mu\nu} &=&  -\eB^2 g^2 \frac{1}{3!} \int \frac{d^4l}{(2 \pi)^4} \int_0^\infty dx_1 dx_2 \ x_2^3 \nonumber \\
 &&\times \exp\Bigl[(x_1+x_2) (l^2-m_f^2) + \frac{x_1 x_2}{x_1+x_2}p^2 \Bigr] \nonumber \\ &&\times \text{Tr}\Big\{\gamma^\nu\gamma^\mu \left(l+\frac{x_1 p}{x_1+x_2}\right)_\perp^2m_f^2\Big\}.
 \label{eqsI1I2I3}
\eea
Taking the trace in Eq.~(\ref{eqsI1I2I3}) and integrating over the internal momenta, discarding odd powers of $l$, and using the tensor basis in Eq.~(\ref{elembasis}) we can write
\bea
I_1^{\mu \nu} &=&2 \left(\frac{i}{16\pi^2}\right) \eB^2 g^2  \int_0^\infty dx_1 dx_2 \ x_2^2 \nonumber \\
 &&\times \exp\Bigl[-(x_1+x_2)m_f^2 + \frac{x_1 x_2}{x_1+x_2}p^2 \Bigr] \nonumber \\
 &\times& \Bigg[ \frac{g_\para^{\mu \nu }}{(x_1+x_2)^3}  +\frac{x_1 x_2}{(x_1+x_2)^4} (p^2\mathcal{P}_0^{\mu \nu} - p_\perp \mathcal{P}_\parallel^{\mu \nu} \nonumber \\
 &+&(p_\parallel^2-2p_\perp^2)\mathcal{P}_\perp^{\mu \nu} - g_\perp^{\mu \nu}p^2)  \Biggr], \nonumber 
 \eea
 \bea
I_2^{\mu \nu} &=&-2\left(\frac{i}{16\pi^2}\right)\eB^2 g^2 \frac{1}{3} \int_0^\infty dx_1 dx_2 \ x_2^3 \nonumber \\
 &&\times \exp\Bigl[-(x_1+x_2) m_f^2 + \frac{x_1 x_2}{x_1+x_2}p^2 \Bigr] \nonumber \\
&&\times \Biggl[  g_\parallel^{\mu \nu} \left(\frac{2}{(x_1+x_2)^4}+\frac{x_1^2}{(x_1+x_2)^5}p_\perp^2\right) \nonumber \\
&-&  g_\perp^{\mu \nu} \left(-\frac{1}{(x_1+x_2)^4}-\frac{x_1^2}{(x_1+x_2)^5}p_\perp^2\right) \nonumber \\
&-&\frac{x_1 x_2}{(x_1+x_2)^2} g^{\mu \nu} p^2 \left(\frac{1}{(x_1+x_2)^3}+\frac{x_1^2}{(x_1+x_2)^4}p_\perp^2\right) \nonumber\\
&&+  \frac{2 x_1 x_2}{(x_1+x_2)^2} p^2 \left( \mathcal{P}_0^{\mu \nu}+ \mathcal{P}_\parallel^{\mu \nu}+\mathcal{P}_\perp^{\mu \nu}  \right) \nonumber \\
&\times& \left(\frac{1}{(x_1+x_2)^3}+\frac{x_1^2}{(x_1+x_2)^4}p_\perp^2\right)    \nonumber \\
&+& \frac{x_1(x_1-x_2)}{(x_1+x_2)^5} ( g_\perp^{\mu \nu}p^2  -p^2 \mathcal{P}_0^{\mu \nu}+p_\perp^2 \mathcal{P}_\parallel^{\mu \nu} \nonumber \\
&&- (p_\para^2-2p_\perp^2)\mathcal{P}_\perp^{\mu \nu} \Biggr], \nonumber \eea
\bea
I_3^{\mu \nu} &=& -2\left(\frac{i}{16\pi^2}\right)\eB^2 g^2 \frac{1}{3} \int_0^\infty dx_1 dx_2 \ x_2^3 \nonumber \\
&&\times \exp\Bigl[-(x_1+x_2)m_f^2 + \frac{x_1 x_2}{x_1+x_2}p^2 \Bigr] \nonumber \\ &&\times g^{\mu\nu} m_f^2 \left(\frac{1}{(x_1+x_2)^3}+ \frac{x_1^2}{(x_1+x_2)^4}p_\perp^2\right).
\eea
Adding up the results for $I_1^{\mu \nu}$, $I_2^{\mu \nu}$, and $I_3^{\mu \nu}$, we can write $\Pi_{(2,0)}^{\mu\nu}$ as
\begin{widetext}
 \bea
 i\Pi_{(2,0)}^{\mu\nu} &=&-2 \left(\frac{i}{16\pi^2}\right) \eB^2 g^2  \int_0^\infty dx_1 dx_2 \exp\Bigl[-(x_1+x_2)m_f^2 + \frac{x_1 x_2}{x_1+x_2}p^2 \Bigr] \nonumber \\
 &\times& \Biggl( -x_2^2\Bigg[ \frac{g_\para^{\mu \nu }}{(x_1+x_2)^3}  +\frac{x_1 x_2}{(x_1+x_2)^4} (p^2\mathcal{P}_0^{\mu \nu}- p_\perp \mathcal{P}_\parallel^{\mu \nu} + (p_\parallel^2-2p_\perp^2)\mathcal{P}_\perp^{\mu \nu} - g_\perp^{\mu \nu}p^2)  \Biggr]  \nonumber \\
 &+&  \frac{x_2^3}{3} \Biggl[ g_\parallel^{\mu \nu} \left(\frac{2}{(x_1+x_2)^4}+\frac{x_1^2}{(x_1+x_2)^5}p_\perp^2\right)-  g_\perp^{\mu \nu} \left(-\frac{1}{(x_1+x_2)^4}-\frac{x_1^2}{(x_1+x_2)^5}p_\perp^2\right) \nonumber \\
&-&\frac{x_1 x_2}{(x_1+x_2)^2} g^{\mu \nu} p^2 \left(\frac{1}{(x_1+x_2)^3}+\frac{x_1^2}{(x_1+x_2)^4}p_\perp^2\right)+  \frac{2 x_1 x_2}{(x_1+x_2)^2} p^2 \left( \mathcal{P}_0^{\mu \nu}+ \mathcal{P}_\parallel^{\mu \nu}+\mathcal{P}_\perp^{\mu \nu}  \right) \nonumber \\
&\times& \left(\frac{1}{(x_1+x_2)^3}+\frac{x_1^2}{(x_1+x_2)^4}p_\perp^2\right)+ \frac{x_1(x_1-x_2)}{(x_1+x_2)^5} ( g_\perp^{\mu \nu}p^2  -p^2 \mathcal{P}_0^{\mu \nu}+p_\perp^2 \mathcal{P}_\parallel^{\mu \nu}- (p_\para^2-2p_\perp^2)\mathcal{P}_\perp^{\mu \nu}) \nonumber\\
&+& g^{\mu\nu} m_f^2 \left(\frac{1}{(x_1+x_2)^3}+ \frac{x_1^2}{(x_1+x_2)^4}p_\perp^2\right)\Biggr] \Biggr).
\label{final20}
\eea
The full expression for the gluon polarization tensor up to ${\mathcal{O}}(B^2)$ is obtained after including the contribution from the third Feynman diagram depicted in Fig.~\ref{fig2} corresponding to the term $\Pi_{(0,2)}^{\mu\nu}$. Notice, however, that this last term can be obtained from Eq.~(\ref{final20}) after the exchange $x_1 \leftrightarrow x_2$. Hence, we immediately get
\bea
i\Pi_{(0,2)}^{\mu\nu} &=&-2 \left(\frac{i}{16\pi^2}\right) \eB^2 g^2  \int_0^\infty dx_1 dx_2 \exp\Bigl[-(x_1+x_2)m_f^2 + \frac{x_1 x_2}{x_1+x_2}p^2 \Bigr] \nonumber \\
 &\times& \Biggl(-x_1^2 \Bigg[ \frac{g_\para^{\mu \nu }}{(x_1+x_2)^3}  +\frac{x_1 x_2}{(x_1+x_2)^4} (p^2\mathcal{P}_0^{\mu \nu} - p_\perp \mathcal{P}_\parallel^{\mu \nu}+(p_\parallel^2-2p_\perp^2)\mathcal{P}_\perp^{\mu \nu} - g_\perp^{\mu \nu}p^2)  \Biggr] \nonumber \\
 &+& \Biggl[ \frac{x_1^3}{3} g_\parallel^{\mu \nu} \left(\frac{2}{(x_1+x_2)^4}+\frac{x_2^2}{(x_1+x_2)^5}p_\perp^2\right)-  g_\perp^{\mu \nu} \left(-\frac{1}{(x_1+x_2)^4}-\frac{x_2^2}{(x_1+x_2)^5}p_\perp^2\right) \nonumber \\
&-&\frac{x_1 x_2}{(x_1+x_2)^2} g^{\mu \nu} p^2 \left(\frac{1}{(x_1+x_2)^3}+\frac{x_2^2}{(x_1+x_2)^4}p_\perp^2\right)+  \frac{2 x_1 x_2}{(x_1+x_2)^2} p^2 \left( \mathcal{P}_0^{\mu \nu}+ \mathcal{P}_\parallel^{\mu \nu}+\mathcal{P}_\perp^{\mu \nu}  \right) \nonumber \\
&\times& \left(\frac{1}{(x_1+x_2)^3}+\frac{x_2^2}{(x_1+x_2)^4}p_\perp^2\right)+ \frac{x_2(x_2-x_1)}{(x_1+x_2)^5} ( g_\perp^{\mu \nu}p^2  -p^2 \mathcal{P}_0^{\mu \nu}+p_\perp^2 \mathcal{P}_\parallel^{\mu \nu} - (p_\para^2-2p_\perp^2)\mathcal{P}_\perp^{\mu \nu} \nonumber\\
&+& g^{\mu\nu} m_f^2 \left(\frac{1}{(x_1+x_2)^3}+ \frac{x_2^2}{(x_1+x_2)^4}p_\perp^2\right)\Biggr] \Biggr).
\label{final02}
\eea
Equations~(\ref{final20}) and (\ref{final02}) can be written together. Introducing the change of variables $x_1=s (1-y)$ and $x_2= s y$, we obtain
\bea 
i\Pi_{(2,0)}^{\mu\nu}&+&i\Pi_{(0,2)}^{\mu\nu} = -\frac{2}{3}\left(\frac{i}{16\pi^2}\right)g^2\eB^2\int_0^1 dy \int_0^\infty  ds  s \exp\Bigl[-s(m_f^2 + y(y-1)p^2) \Bigr]\nonumber\\
&\times&\Bigg\{ g_\para^{\mu\nu} \Biggl[-\frac{1}{ s} + s y^3 (y-1)^3 p_\parallel^2 p_\perp^2 - s y^3 (y-1)^3 p_\perp^4 - y (y-1) (1+ 2y (y-1))p_\perp^2   \nonumber \\
&& +  y (y-1) (1+ 2y (y-1)) p_\parallel^2 + m_f^2 \left( 1- 3y (y-1) +  s y^2 (y-1)^2 p_\perp^2 \right) \Biggl]\nonumber \\
&&+\ g_\perp^{\mu\nu} \Biggl[ \frac{1+ 3y (y-1)}{s }+ s y^3 (y-1)^3  p_\parallel^2 p_\perp^2 - s y^3 (y-1)^3  p_\perp^4 + y (y-1) p_\perp^2 + y (1+ y^2(y-2)) p_\parallel^2 \nonumber \\
&&+ m_f^2 \left( 1- 3y (y-1) + s y^2 (y-1)^2 p_\perp^2 \right)   \Biggr]+\mathcal{P}_0^{\mu\nu}\left[ -4 y^2 (y-1)^2 p^2 - 2sy^3 (y-1)^3  p^2 p_\perp^2  \right]  \nonumber \\
&&+\mathcal{P}_\para^{\mu\nu}\Bigl[ - 2 y(y-1)(1+3y (y-1)) p_\parallel^2 +  4 y^2 (y-1)^2 p_\perp^2 -2 s y^3 (y-1)^3 p^2 p_\perp^2  \Bigr] \nonumber \\
&&+\mathcal{P}_\perp^{\mu\nu}\Bigl[-  4 y^2 (y-1)^2 p_\parallel^2 +  2 y (1+ y^2(y-2))p_\perp^2 -2 s y^3 (y-1)^3 p^2 p_\perp^2 \Bigr] \Bigg\}. 
\eea 


Adding the contributions from $\Pi_{(1,1)}^{\mu\nu}$, $\Pi_{(2,0)}^{\mu\nu}$, and  $\Pi_{(0,2)}^{\mu\nu}$, we finally obtain
\bea 
i\Pi^{\mu\nu}&=& -\frac{2}{3}\left(\frac{i}{16\pi^2}\right)g^2\eB^2\int_0^1 dy \int_0^\infty  ds  s  \exp\Bigl[-s(m_f^2 + y(y-1)p^2) \Bigr]\nonumber\\
&\times&\Bigg\{ g_\para^{\mu\nu} \Biggl[-\frac{1}{ s} + s y^3 (y-1)^3 p_\parallel^2 p_\perp^2  - s y^3 (y-1)^3 p_\perp^4 - y (y-1) (1+ 2y (y-1))p_\perp^2   \nonumber \\
&& +  y (y-1) p_\parallel^2 + m_f^2 \left( 1 +  s y^2 (y-1)^2 p_\perp^2 \right) \Biggl]\nonumber \\
&&+\ g_\perp^{\mu\nu} \Biggl[ \frac{1+6y(y-1) }{s }+ s y^3 (y-1)^3  p_\parallel^2 p_\perp^2 - s y^3 (y-1)^3  p_\perp^4 + y (y-1) p_\perp^2
\nonumber\\
&&\ - y(y-1) (1+ 2y(y-1)) p_\parallel^2 + m_f^2 \left( 1+6y(y-1) + s y^2 (y-1)^2 p_\perp^2 \right)   \Biggr] \nonumber \\ &&+\mathcal{P}_0^{\mu\nu}\left[ -4 y^2 (y-1)^2 p^2 - 2sy^3 (y-1)^3  p^2 p_\perp^2  \right]  \nonumber \\
&&+\mathcal{P}_\para^{\mu\nu}\left[ -2y (y-1) p_\parallel^2 +  4 y^2 (y-1)^2 p_\perp^2 -2 s y^3 (y-1)^3 p^2 p_\perp^2  \right] \nonumber \\
&&+\mathcal{P}_\perp^{\mu\nu}\left[-  4 y^2 (y-1)^2 p_\parallel^2 +  2 y (1+ y^2(y-2))p_\perp^2 -2 s y^3 (y-1)^3 p^2 p_\perp^2 \right] \Bigg\}. 
\label{piwithA}
\eea 

Notice that Eq.~(\ref{piwithA}) contains terms proportional to the tensors $g_\parallel^{\mu \nu}$ and $g_\perp^{\mu \nu}$. Their coefficients are the $A_1$ and  $A_2$ factors in Eq.~(\ref{eq24}), respectively. Nevertheless, it is easy to show that the coefficient $A_1$ vanishes when carrying out the integration over $s$. In a similar fashion, $A_2$ also vanishes upon integration over both $s$ and $y$. Therefore we can finally write
\bea 
i\Pi^{\mu\nu}&=& -\frac{2}{3}\left(\frac{i}{16\pi^2}\right)g^2\eB^2\int_0^1 dy \int_0^\infty  ds\,s  \exp\Bigl[-s(m_f^2 + y(y-1)p^2) \Bigr]\nonumber\\
&\times&\Bigg\{ \mathcal{P}_0^{\mu\nu}\left[ -4 y^2 (y-1)^2 p^2 - 2sy^3 (y-1)^3  p^2 p_\perp^2  \right]  \nonumber \\
&&+\mathcal{P}_\para^{\mu\nu}\left[ -2y (y-1) p_\parallel^2 +  4 y^2 (y-1)^2 p_\perp^2 -2 s y^3 (y-1)^3 p^2 p_\perp^2  \right] \nonumber \\
&&+\mathcal{P}_\perp^{\mu\nu}\left[-  4 y^2 (y-1)^2 p_\parallel^2 +  2 y (1+ y^2(y-2))p_\perp^2 -2 s y^3 (y-1)^3 p^2 p_\perp^2 \right] \Bigg\}.
\label{finalPi}
\eea 
For the momentum range $0\leq p^2\leq 4m_f^2$, Eq.~(\ref{finalPi}) lends itself to perform the integrations analytically. Thus, we finally obtain the explicit  coefficients of the three basis tensor structures shown in Eq.~(\ref{basistensor}). These are given by
%
%
%
\begin{subequations}
\bea
P_0&=&\frac{i g^2\eB^2}{6\pi^2}\Bigg \{ \frac{p^2-6 m^2}{p^2 \left(p^2-4 m^2\right)}+\frac{\left(p^2-10 m^2\right) \left(p^2-3 m^2\right)}{p^4 \left(p^2-4 m^2\right)^2}\pt^2 \nn\\
&+& \left[\frac{8 m^2 \left(p^2-3 m^2\right) }{(p^2)^{3/2} \left(4 m^2-p^2\right)^{3/2}}-\frac{12 m^2 \left(10 m^4+(p^2-6m^2)p^2\right) }{\left(4m^2-p^2\right)^{5/2} (p^2)^{5/2}}\pt^2\right]\arctan\left(\sqrt{\frac{p^2}{4m_f^2-p^2}}\right) \Bigg \},
\label{coefP0}
\eea
\bea
P_\parallel&=&\frac{i g^2\eB^2}{6\pi^2}\Bigg \{ -\frac{\pp^2}{\left(4 m^2-p^2\right) p^2}-\frac{(p^2-6 m^2)\pt^2}{p^4 \left(p^2-4 m^2\right)}+\frac{\left(p^2-10 m^2\right) \left(p^2-3 m^2\right)\pt^2}{p^4 \left(p^2-4 m^2\right)^2} \nn\\
&+&\left[\frac{-2 \left(p^2-2m^2\right) \pp^2}{\left(4 m^2-p^2\right)^{3/2} (p^2)^{3/2}}-\frac{8 m^2 \left(p^2-3m^2\right)\pt^2}{\left(4 m^2-p^2\right)^{3/2} (p^2)^{5/2}}-\frac{12 m^2 \left(10 m^4+(p^2-6m^2)p^2\right)}{\left(4 m^2-p^2\right)^{5/2} (p^2)^{5/2}}\pt^2\right]\arctan\left(\sqrt{\frac{p^2}{4m_f^2-p^2}}\right) \Bigg \}, \nn\\
\label{coefPL}
\eea
\bea
P_\perp&=&\frac{i g^2\eB^2}{6\pi^2}\Bigg \{ \frac{\left(p^2-6 m^2\right)\pp^2}{p^4 \left(p^2-4 m^2\right)}-\frac{\left(6 m^2+p^2\right)\pt^2}{2\left(4 m^2-p^2\right) p^4}+\frac{\left(p^2-10 m^2\right) \left(p^2-3 m^2\right)\pt^2}{p^4 \left(p^2-4 m^2\right)^2} \nn\\
&+&\left[\frac{8 m^2 \left(p^2-3m^2\right)\pp^2}{\left(4 m^2-p^2\right)^{3/2} (p^2)^{5/2}}+\frac{2 \left(6 m^4-p^4\right)\pt^2}{\left(4 m^2-p^2\right)^{3/2} (p^2)^{5/2}}-\frac{12 m^2 \left(10 m^4+(p^2-6m^2)p^2\right)}{\left(4 m^2-p^2\right)^{5/2} (p^2)^{5/2}}\pt^2\right]\arctan\left(\sqrt{\frac{p^2}{4m_f^2-p^2}}\right) \Bigg \}. \nn \\
\label{coefPT}
\eea
\end{subequations}

\end{widetext}
It is worth noticing that the result in Eqs.~(\ref{coefP0}), (\ref{coefPL}) and (\ref{coefPT}) coincide with the ${\mathcal{O}}(B^2)$ found in Ref.~\cite{polgen}. 

\section{Gluon dispersion relation and polarizations}\label{secIV}

\begin{figure}[t!]
    \centering
    \includegraphics[scale=0.35]{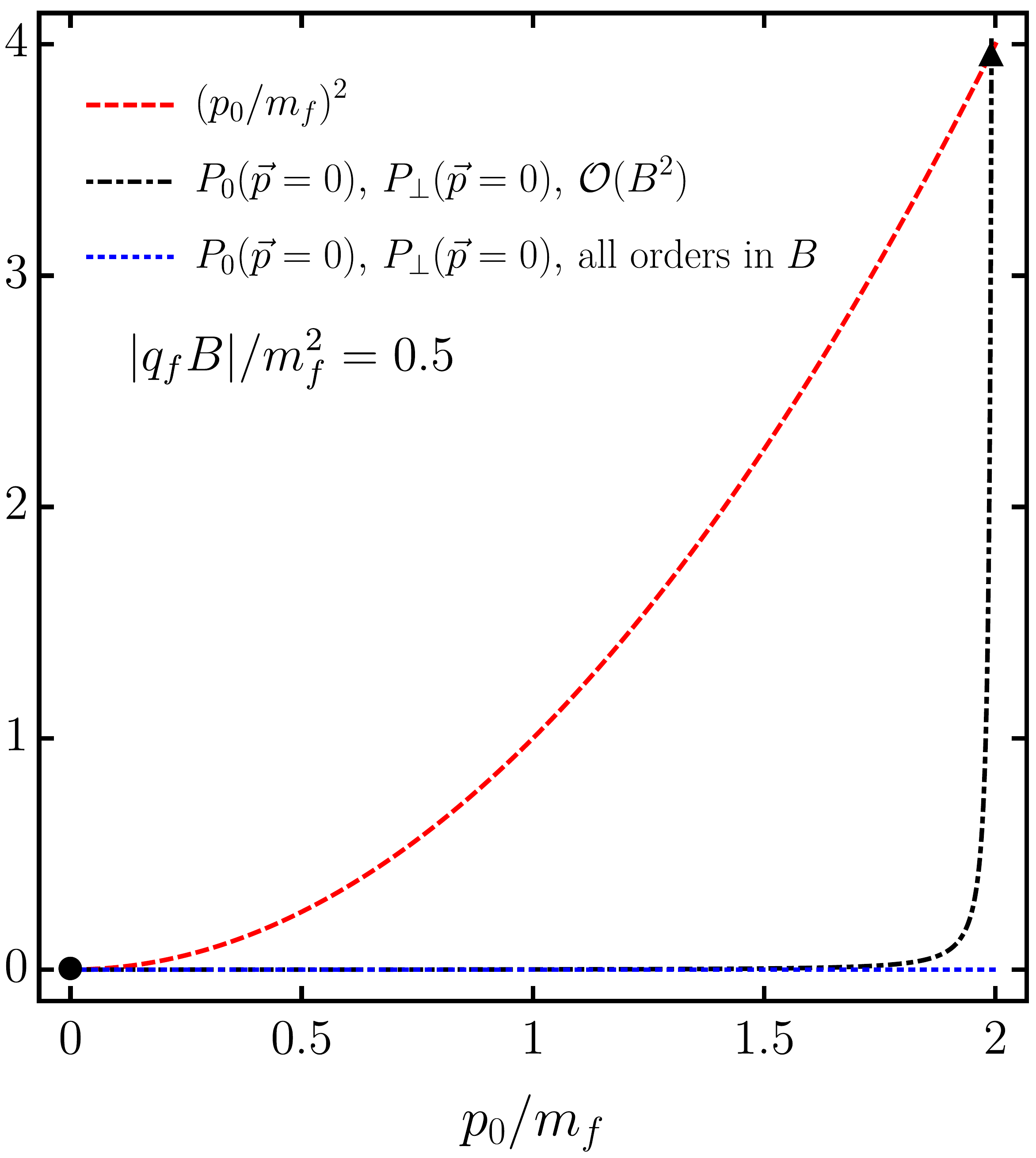}
    \caption{The gluon magnetic mass obtained from the coefficients $P_0$ and $P_\perp$ found as the intersection of the curves corresponding to the
    expressions on the left- and right-hand sides of Eq.~(\ref{solint}). A spurious solution with a finite gluon mass is obtained when
    computing the coefficients at ${\mathcal{O}}(B^2)$ (indicated by the full triangle). However, the numerical calculation
    to all orders in $B$ shows that only the solution $p_0=0$ (indicated by the full circle) remains.}
    \label{Fig:debye_mass}
\end{figure}

The dispersion properties and polarization strength of the propagating gluon modes are encoded in the coefficients of the tensor structures in Eq.~(\ref{finalPi}).  In order to extract these properties, we proceed first to find the possible magnetic field-induced modification to the gluon mass and dispersion relations. For this purpose, henceforth, we use $\alpha_s=g^2/4\pi=0.3$ in the analysis. Figure~\ref{Fig:debye_mass} shows plots representing the solutions of the equation 
\bea
p_0^2=P_i(p_0)\Big{|}_{p_\perp=0,p_3=0}
\label{solint}
\eea
for $i=\left(0, 
\perp\right)$ as functions of the ratio
$p_0/m_f$ and for a fixed value of the ratio $|q_fB|/m_f^2=0.5$.
The intercept of the curves on the left- and right-hand side of
Eq.~(\ref{solint}) corresponds to the gluon magnetic mass for each
mode. 
The figure shows the case where
these coefficients are computed up
to ${\mathcal{O}}(B^2)$ as well as to all orders in $B$. Notice that
for the former, in addition to the solution at $p_0=0$, there is a
second intercept that would correspond to a finite gluon mass for
both modes. Nevertheless, this solution is spurious, as can be seen
from the numerical computation of the right-hand side of Eq.~(\ref{solint})
to all orders in $B$, using the findings of Ref.~\cite{polgen}, where
the second intercept  disappears. In contrast, the curve for the
right-hand side of Eq.~(\ref{solint}) up to $\mathcal{O}(B^2)$ for
the case of $P_\para$ (not shown in the figure) is found to become asymptotically negative as $p_0/m_f$ approaches 2, never intersecting the parabola on the left-hand side of Eq.~(\ref{solint}). Consequently, a spurious
solution does not exists for this mode.\par
%
%
A similar feature is found for the dispersion relation of each mode. 
Due to the loss of Lorentz invariance, the coefficients $P_i$ are in
principle functions of the two variables $p_\para^2$ and $p_\perp^2$
and not of $p_0^2$ and $|\vec{p}\,|^2=p_\perp^2+p_3^2$, which could
cast doubts on whether or not the gluons move along the light cone,
as should correspond to a massless excitation. In order to check if
this is the case, we parametrize the relation between $p_3^2$ and
$p_\perp^2$ as $p_3^2=a\, p_\perp^2$. Written in this fashion, the
case $a=0$ corresponds to motion in the transverse plane, whereas the
case $a\to\infty$ corresponds to motion along the magnetic field
direction. 
%
\begin{figure}[t!]
    \centering
    \includegraphics[scale=0.557]{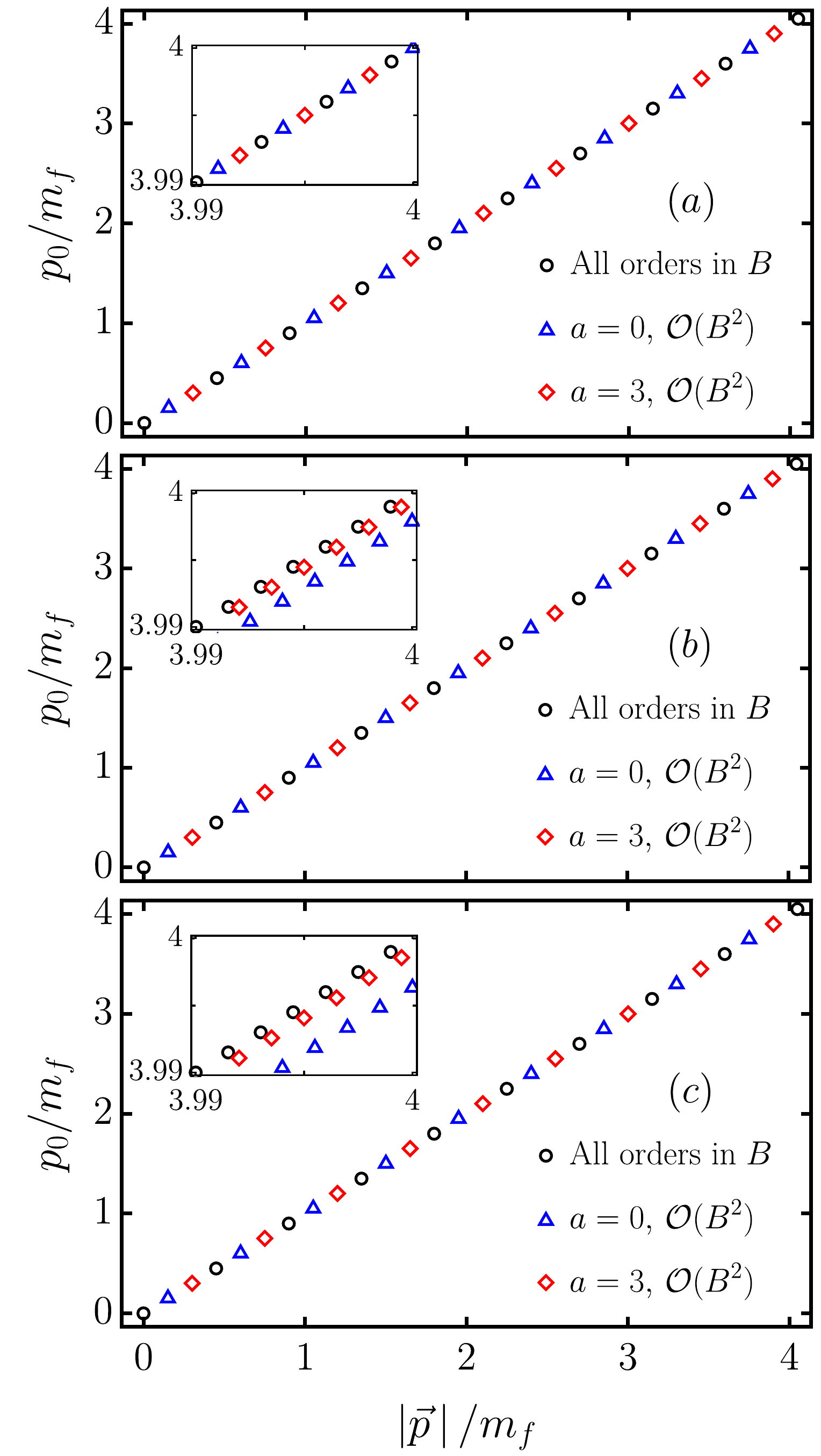}
    \caption{Dispersion relation obtained for the modes corresponding to 
    (a) $P_0$, (b) $P_\para$, and (c) $P_\perp$ for two different values of the parameter $a$ computed up to ${\mathcal{O}}(B^2)$ and to all orders in $B$ for $|q_fB|/m_f^2=0.5$. In each case only
    the non-spurious solution of Eq.~(\ref{solint}) is considered. The calculation to all orders corresponds to motion along
    the light cone for all three modes. In the case of (a), the approximation
    provides the exact solution. For (b) and (c), slight deviations
    with respect to the exact result are found and the light cone is approached to from below.}
    \label{Fig:dispersion_relation}
\end{figure}
%
Figure~\ref{Fig:dispersion_relation} shows the dispersion
relation for each mode for a given value of the ratio $|q_fB|/m_f^2$ and two values of
$a$ as a function of $|\vec{p}|=\sqrt{(1+a)p_\perp^2}$ up to ${\mathcal{O}}(B^2)$,
using the coefficients found in Eq.~(\ref{finalPi}), and for the case of the
calculation to all orders in $B$, using the findings of Ref.~\cite{polgen}.
Notice also that the calculation to all orders in $B$ corresponds to a  dispersion  relation
representing motion along the light cone, whereas the calculation up
to ${\mathcal{O}}(B^2)$ deviates little from light cone propagation.
Just as for the case of the gluon mass, there is a second spurious
solution for the modes corresponding to $P_0$ and $P_\perp$ that disappears when the calculation is performed to all
orders in $B$. 
%
\begin{figure}[t!]
    \centering
    \includegraphics[scale=0.566]{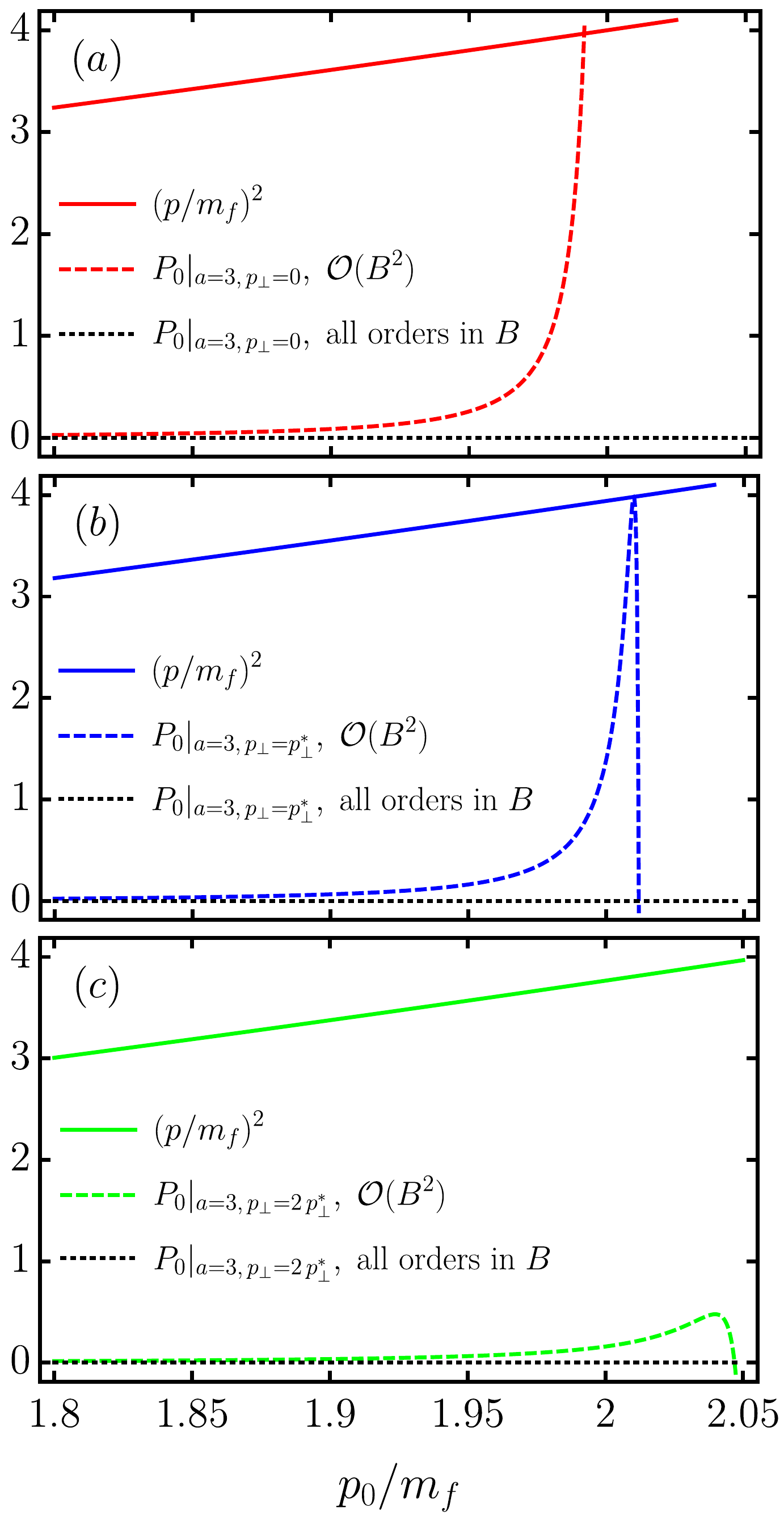}
    \caption{Solutions for the dispersion relation for the mode $i=0$ as 
    functions of $p_0/m_f$ for
    a fixed value of $a$ and three different values of $p_\perp$. The
    plots show the results of the calculation up to ${\mathcal{O}}(B^2)$ and
    to all orders in $B$ represented as the intercept between the solid and
    dashed or dotted lines.
    In the case of the calculation at ${\mathcal{O}}(B^2)$ and for $p_\perp\leq p_\perp^*$, solutions
    other than the light cone arise but these disappear for $p_\perp>p_\perp^*$, where
    $p_\perp^*=0.1209 \ m_f$. For the computation to
    all orders in $B$, none of these spurious solutions appear.}
    \label{Fig:dispersion_relation_abc}
\end{figure}
%
This is illustrated in Fig.~\ref{Fig:dispersion_relation_abc}
for $P_0$, which shows a second intercept between the left- and right-hand sides
of Eq.~(\ref{solint}) when the latter is computed using the coefficient
found in Eq.~(\ref{coefP0}) for three different values of $p_\perp$. 
The spurious solution develops for $p_\perp<p_\perp^*$, where
$p_\perp^*=0.1209 \ m_f$; above this value, only the non-spurious solution exists.
As in the case of the gluon magnetic mass,
this behavior is also found for $P_\perp$ but it is nonexistent for $P_\para$. 
The message that can be drawn from the above analysis is that, as far
as the propagation properties of the gluons are concerned, the
approximation for the gluon polarization tensor to lowest non-trivial
order in $B$ is reliable provided the second (spurious) solution,
corresponding to a finite gluon mass, is discarded.
%
\begin{figure}[h!]
    \centering
    \includegraphics[scale=0.623]{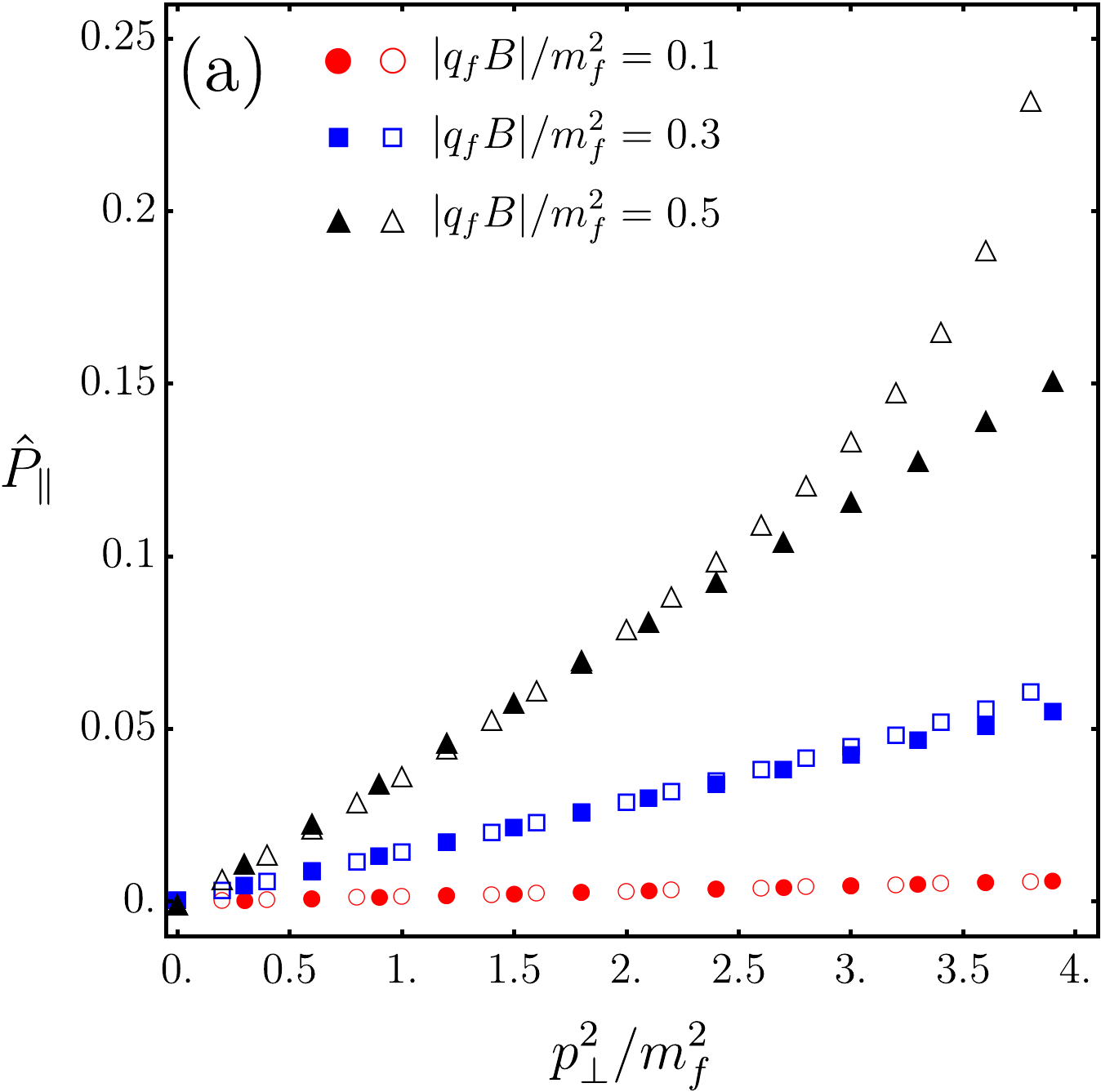}\\
    \vspace{0.5cm}
    \includegraphics[scale=0.623]{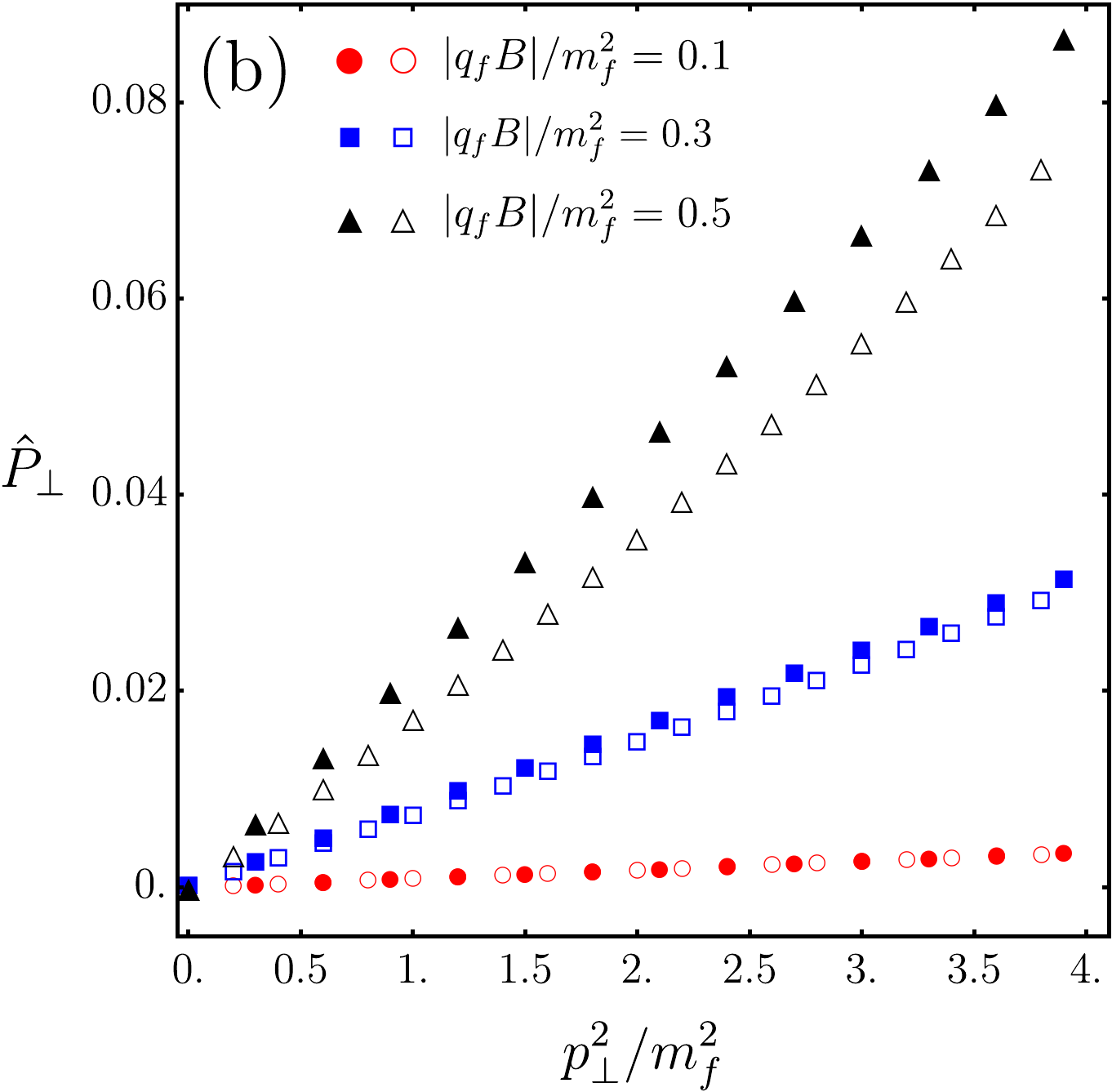}
    \caption{Normalized tensor coefficients $\hat{P}_{\para,\perp}$ as a function of $\pt^2/m_f^2$ computed from the weak field expansion given by Eqs.~(\ref{coefPL})-(\ref{coefPT}) (full symbols) and compared with the general expressions found in Ref.~\cite{polgen} (open symbols). The results are shown for the three values of the ratio $\eB/m_f^2=0.1,0.3$ and $0.5$ and with the on-shell condition $p^2=0$.}
    \label{Fig:Pl_and_PT}
\end{figure}

In order to study the strength of the polarization modes for real gluons, we now proceed to analyse the behavior of Eqs.~(\ref{coefP0}), (\ref{coefPL}) and (\ref{coefPT}) when $p_0^2=|\vec{p\,}|$. Notice that for real gluons $p_\parallel^2\to p_\perp^2$ and therefore Eqs.~(\ref{coefP0}), (\ref{coefPL}), and (\ref{coefPT}) become functions of only $p_\perp^2$ and also that  Eq.~(\ref{coefP0}) vanishes for real gluons with $p^2=0$. Figures~\ref{Fig:Pl_and_PT} (a) and (b) show the normalized coefficients $\hat{P}_{\para}\equiv\left(8\pi^2/g^2m_f^2\right)P_{\para}$ and $\hat{P}_{\perp}\equiv\left(8\pi^2/g^2m_f^2\right)P_{\perp}$ as functions of $p_\perp^2/m_f^2$ for three values of the ratio $|q_fB|/m_f^2$ with the on-shell condition. Also, the horizontal axis is restricted to the range $0\leq\pt^2/m_f^2\leq 4$, where the weak field expressions in Eqs.~(\ref{coefP0})--(\ref{coefPT}) are valid. The full symbols correspond to the results obtained from Eqs.~(\ref{coefPL}) and (\ref{coefPT}), whereas the open symbols come from the general expressions given in Ref.~\cite{polgen}. Notice that the weak field expansion is accurate for small values of $\eB/m_f^2$.

\section{Summary and conclusions}\label{concl}

In this work we have computed and analysed the properties of the gluon polarization tensor in the presence of a weak, uniform, and constant magnetic field. For this purpose, we derive the fermion propagator for the case when the magnetic field is the smallest energy (squared) scale. We showed two different and alternative methods to derive the propagator. The first one starts from the Schwinger's proper-time representation. In this case we find that it is enough to perform a Taylor expansion to then integrate over the proper time parameter. The second method starts from the propagator written in terms of the Landau Levels representation. This method shows how one can replace the sum over Landau Levels by a series in powers of the magnetic field strength. The condition to be satisfied is that $|q_fB|<m_f^2$. Both methods coincide and the result is illustrated writing the propagator up to $\mathcal{O}(B
^6)$.
The expression found in this work for the fermion propagator is in complete agreement with the one reported in Ref.~\cite{Chyi}.
\par
With the expression for the fermion propagator in the weak field approximation at hand, we compute the first non trivial magnetic contribution to the polarization tensor, namely, the $\mathcal{O}(B^2)$ contribution. 
Since the magnetic field breaks Lorentz symmetry, the polarization tensor is expressed in terms of three tensor structures, $\mathcal{P}^{\mu \nu}_\parallel$, $\mathcal{P}^{\mu \nu}_\perp$, and $\mathcal{P}^{\mu \nu}_0$.
When the computation is explicitly carried out, two other non-physical tensor structures appear, $g^{\mu \nu}_\parallel$ and $g^{\mu \nu}_\perp$, whose coefficients are shown to vanish. The coefficients of the three physical tensor structures, $P_0$, $P_\parallel$, and $P_\perp$, are provided explicitly 
and they coincide with the result at $\mathcal{O}(B^2)$ from the expansion to the same order starting from the expression valid to all orders,  reported in Ref.~\cite{polgen}. The agreement between the results in this work and those in  Ref.~\cite{polgen} shows that the quantum corrections due to the magnetic field, in the weak field approximation are reliable as the ratio $|q_fB|/m_f^2$ becomes smaller. 
\par
We have also studied the dispersion properties and the polarization strength of the propagating gluon modes. Since, the magnetic field breaks Lorentz symmetry, the gluon four-momentum splits into  $p_\parallel$ and $p_\perp$ components and thus the polarization tensors become themselves functions of these components. Therefore, it is not obvious that gluons do not develop a magnetic mass and that their dispersion  relation is not modified. As we have shown, this is indeed the case and gluons move along the light cone even in the presence of a background magnetic field. This is confirmed by comparing the results at ${\mathcal{O}}(B^2)$ with the computation performed at all orders using the results from Ref.~\cite{polgen}. The coincidence of the results happens provided a second, spurious solution for the mass and dispersion relations for the $P_0$ and $P_\perp$ coefficients, is ignored.
\par
The findings of this work are now suited to be extended to the finite temperature and density case when either one is larger than the field strength or situations where the QCD background consists of a large occupation number, as is the case of the Color Glass Condensate~\cite{GluonHelicity}. This is work currently being developed and that will be reported elsewhere. 

\section*{Acknowledgements}
This work was supported in part by UNAM-DGAPA-PAPIIT grant number IG100219. L. A. H. acknowledges support from a PAPIIT-DGAPA-UNAM fellowship. R. Zamora acknowledges support from FONDECYT (Chile) under grant No. 1200483.

\appendix
\section{Derivation of Eqs.~(\ref{iSfinalfromLandau}) and~(\ref{Gdefinitions})}\label{ApDerivationsFromLandau}
From Eq.~(\ref{eq:MiranskyPropagator}) three sums can be identified:
\begin{subequations}
\bea
s_1&=&2ie^{-\alpha}\sum_{n=0}^\infty(-1)^n\frac{L_n^0(2\alpha)}{\pp^2-m_f^2-2n\eB}\nn\\
&=&\frac{2ie^{-\alpha}}{\pp^2-m_f^2}\sum_{n=0}^\infty(-1)^n\frac{L_n^0(2\alpha)}{1-\frac{2n\eB}{\pp^2-m_f^2}},
\label{S1}
\eea
\bea
s_2&=&-2ie^{-\alpha}\sum_{n=1}^\infty(-1)^n\frac{L_{n-1}^0(2\alpha)}{\pp^2-m_f^2-2n\eB}\nn\\
&=&-\frac{2ie^{-\alpha}}{\pp^2-m_f^2}\sum_{n=1}^\infty(-1)^n\frac{L_{n-1}^0(2\alpha)}{1-\frac{2n\eB}{\pp^2-m_f^2}},
\label{S2}
\eea
\bea
s_3&=&4ie^{-\alpha}\sum_{n=1}^\infty(-1)^n\frac{L_{n-1}^1(2\alpha)}{\pp^2-m_f^2-2n\eB}\nn\\
&=&\frac{4ie^{-\alpha}}{\pp^2-m_f^2}\sum_{n=1}^\infty(-1)^n\frac{L_{n-1}^1(2\alpha)}{1-\frac{2n\eB}{\pp^2-m_f^2}},
\label{S3}
\eea
\label{sums}
\end{subequations}
with $\alpha=\pt^2/\eB$. 

The assumption that the magnetic field is the weakest scale allows us to promote the denominator of Eqs.~(\ref{sums}) to a geometric series, as it was shown in Eq.~(\ref{DenominatortoGeomSum}), which yields two nested sums. The sum in $n$ can be performed using the Eqs.~(\ref{Identities1def}) and~(\ref{DerivativeIdentities1}) so that
\begin{subequations}
\bea
&&s_1=\frac{i}{\pp^2-m_f^2}\sigma_1,
\eea
\bea
&&s_2=\frac{i}{\pp^2-m_f^2}\sigma_2,
\eea
and
\bea
&&s_3=-\frac{i}{\pp^2-m_f^2}\sigma_3,
\eea
\label{s_and_sigma}
\end{subequations}
where
\begin{subequations}
\bea
\sigma_1\equiv\lim_{v\rightarrow0}\sum_{k=0}^{\infty}\left(\frac{i\eB}{\pp^2-m_f^2}\right)^k\frac{\partial^k}{\partial v^k}\left(\frac{e^{iv}}{\cos v}e^{-i\alpha\tan v}\right),\nn\\
\eea
\bea
\sigma_2\equiv\lim_{v\rightarrow0}\sum_{k=0}^{\infty}\left(\frac{i\eB}{\pp^2-m_f^2}\right)^k\frac{\partial^k}{\partial v^k}\left(\frac{e^{-iv}}{\cos v}e^{-i\alpha\tan v}\right),\nn\\
\eea
and
\bea
\sigma_3\equiv\lim_{v\rightarrow0}\sum_{k=0}^\infty\left(\frac{i\eB}{\pp^2-m_f^2}\right)^k\frac{\partial^k}{\partial v^k}\left(\frac{e^{-i\alpha\tan v}}{\cos^2v}\right). \nn\\
\eea

\end{subequations}

The remaining sums cannot be expressed in closed-form. However, it is possible to find the contribution at the desired order in the parameter space $(\pp^2,\pt^2,\eB,m_f^2)$. By expanding the sum over $k$ and together with the definitions $x=\pt^2/(\pp^2-m_f^2)$ and $\mathfrak{B}=\eB/(\pp^2-m_f^2),$ it is straightforward to find the following series
\begin{widetext}
\begin{subequations}
\bea
\sigma_1&=&\left(1+x+x^2+x^3+x^4+x^5+x^6+x^7+x^8+x^9+x^{10}+\cdots\right)-\left(1+2 x+3 x^2+4 x^3+5 x^4+6 x^5+7x^6\right.\nn\\
&+&\left.8x^7+9 x^8+10 x^9+11x^{10}+\cdots\right)\mathfrak{B}-2x\left(1+4 x+10 x^2+20 x^3+35 x^4+56 x^5+84 x^6+120 x^7+165 x^8\right.\nn\\
&+&\left.220 x^9+286x^{10}+\cdots\right)\mathfrak{B}^2+2\left(1+8 x+30 x^2+80 x^3+175x^4+336 x^5+588 x^6+960 x^7+1485x^8+2200 x^9\right.\nn\\
&+&\left.3146 x^{10}+\cdots\right)\mathfrak{B}^3+8x\left(2+17 x+77 x^2+252 x^3+672 x^4+1554 x^5+3234 x^6+6204x^7+11154 x^8+19019 x^9\right.\nn\\
&+&\left.31031 x^{10}+\cdots\right)\mathfrak{B}^4-8\left(2+34 x+231 x^2+1008 x^3+3360 x^4+9324 x^5+22638 x^6+49632x^7+100386 x^8\right.\nn\\
&+&\left.190190 x^9+341341 x^{10}+\cdots\right)\mathfrak{B}^5-16x\left(17+248 x+1760 x^2+8480 x^3+31790 x^4+99704 x^5+273416x^6\right.\nn\\
&+&\left.674960 x^7+1530815 x^8+3237520 x^9+6456736x^{10}+\cdots\right)\mathfrak{B}^6+\Op\left(\mathfrak{B}^7\right),
\eea
\bea
\sigma_2&=&\left(1+x+x^2+x^3+x^4+x^5+x^6+x^7+x^8+x^9+x^{10}+\cdots\right)+\left(1+2 x+3 x^2+4 x^3+5 x^4+6 x^5+7x^6\right.\nn\\
&+&\left.8x^7+9 x^8+10 x^9+11x^{10}+\cdots\right)\mathfrak{B}-2x\left(1+4 x+10 x^2+20 x^3+35 x^4+56 x^5+84 x^6+120 x^7+165 x^8\right.\nn\\
&+&\left.220 x^9+286x^{10}+\cdots\right)\mathfrak{B}^2-2\left(1+8 x+30 x^2+80 x^3+175x^4+336 x^5+588 x^6+960 x^7+1485x^8+2200 x^9\right.\nn\\
&+&\left.3146 x^{10}+\cdots\right)\mathfrak{B}^3+8x\left(2+17 x+77 x^2+252 x^3+672 x^4+1554 x^5+3234 x^6+6204x^7+11154 x^8+19019 x^9\right.\nn\\
&+&\left.31031 x^{10}+\cdots\right)\mathfrak{B}^4+8\left(2+34 x+231 x^2+1008 x^3+3360 x^4+9324 x^5+22638 x^6+49632x^7+100386 x^8\right.\nn\\
&+&\left.190190 x^9+341341 x^{10}+\cdots\right)\mathfrak{B}^5-16x\left(17+248 x+1760 x^2+8480 x^3+31790 x^4+99704 x^5+273416x^6\right.\nn\\
&+&\left.674960 x^7+1530815 x^8+3237520 x^9+6456736x^{10}+\cdots\right)\mathfrak{B}^6+\Op\left(\mathfrak{B}^7\right),
\eea
%
and
\bea
\sigma_3&=&\left(1+x+x^2+x^3+x^4+x^5+x^6+x^7+x^8+x^9+x^{10}+\cdots\right)-2 \left(1+4 x+10 x^2+20 x^3+35 x^4+56 x^5+84 x^6\right.\nn\\
&+&\left.120 x^7+165 x^8+220x^9+286 x^{10}+\cdots\right)\Bf^2+8 \left(2+17 x+77 x^2+252 x^3+672 x^4+1554 x^5+3234 x^6+6204x^7\right.\nn\\
&+&\left.11154 x^8+19019 x^9+31031 x^{10}+\cdots\right)\Bf^4-16 \left(17+248 x+1760 x^2+8480 x^3+31790 x^4+99704 x^5+273416x^6\right.\nn\\
&+&\left.674960 x^7+1530815 x^8+3237520 x^9+6456736 x^{10}\right)\Bf^6+\Op\left(\Bf^8\right).
\eea
\label{isseries}
\end{subequations}
\end{widetext}

The power series of $x$ in Eqs.~(\ref{isseries}) can be easily recognized yielding
\begin{subequations}
\bea
\sigma_1&=&\frac{1}{1-x}-\frac{\Bf}{(1-x)^2}-\frac{2x\Bf^2}{(1-x)^4}+\frac{2 (3 x+1)\Bf^3}{(1-x)^5}\nn\\
&+&\frac{8 x (3 x+2)\Bf^4}{(1-x)^7}-\frac{8 \left(15 x^2+18 x+2\right)\Bf^5}{(1-x)^8}\nn\\
&-&\frac{16 x \left(45 x^2+78 x+17\right)\Bf^6}{(1-x)^{10}}+\Op\left(\Bf^7\right),
\eea
\bea
\sigma_2&=&\frac{1}{1-x}+\frac{\Bf}{(1-x)^2}-\frac{2x\Bf^2}{(1-x)^4}-\frac{2 (3 x+1)\Bf^3}{(1-x)^5}\nn\\
&+&\frac{8 x (3 x+2)\Bf^4}{(1-x)^7}+\frac{8 \left(15 x^2+18 x+2\right)\Bf^5}{(1-x)^8}\nn\\
&-&\frac{16 x \left(45 x^2+78 x+17\right)\Bf^6}{(1-x)^{10}}+\Op\left(\Bf^7\right),
\eea
and
\bea
\sigma_3&=&\frac{1}{1-x}-\frac{2\Bf^2}{(1-x)^4}+\frac{8 (3 x+2)\Bf^4}{(1-x)^7}\nn\\
&-&\frac{16 \left(45 x^2+78 x+17\right)\Bf^6}{(1-x)^{10}}+\Op\left(\Bf^8\right).
\eea
\label{s_en_x}
\end{subequations}

The final result is obtained by replacing Eqs.~(\ref{s_en_x}) into Eqs.~(\ref{s_and_sigma}) and Eqs.~(\ref{sums}) in such a way that the matrix structures are collected as
\begin{subequations}
\bea
\slashed{p}=\slashed{p}_\parallel-\slashed{p}_\perp,
\eea
\bea
\Op^+ +\Op^-=1,
\eea
and
\bea
\Op^+ -\Op^-=i\,\sign{q_fB}\gamma^1\gamma^2,
\eea
\end{subequations}
~\\
\noindent
so that at order $\Bf^6$ the fermion propagator is
\bea
&&\!\!\!\!\!\!\!\!\!\!iS(p)=\frac{i}{\pp^2-m_f^2}\Bigg{\{}\frac{1}{1-x}\left(\ps+m_f\right)\nn\\
&+&\left[\frac{\Bf}{(1-x)^2}-\frac{2 (3 x+1)\Bf^3}{(1-x)^5}+\frac{8 \left(15 x^2+18 x+2\right)\Bf^5}{(1-x)^8}\right]\nn\\
&\times&i\,\sign{q_f B}\gamma^1\gamma^2(\ps_\parallel+m_f)\nn\\
&-&\left[\frac{2\Bf^2}{(1-x)^4}-\frac{8 (3 x+2)\Bf^4}{(1-x)^7}-\frac{16 \left(45 x^2+78 x+17\right)\Bf^6}{(1-x)^{10}}\right]\nn\\
&\times&\left(x(\ps_\parallel+m_f)-\slashed{p}_\perp\right)\Bigg{\}},
\eea
which after substituting $x$ yields
\begin{widetext}
\bea
iS(p)=\frac{i\left(\ps+m_f\right)}{p^2-m_f^2}-\mathcal{G}_1(p,B)\,\sign{q_f B}\gamma^1\gamma^2(\ps_\parallel+m_f)-2i(\pp^2-m_f^2)\,\mathcal{G}_2(p,B)\left[\frac{\pt^2}{\pp^2-m_f^2}(\ps_\parallel+m_f)-\slashed{p}_\perp\right],
\label{iSfinalfromLandauApp}
\eea
where
\begin{subequations}
\bea
\mathcal{G}_1(p,B)\equiv\frac{\eB}{(p^2-m_f^2)^2}-2\frac{3\pt^2+\pp^2-m_f^2}{(p^2-m_f^2)^5}\eB^3+8\frac{15\pt^4+18\pt^2(\pp^2-m_f^2)+2(\pp^2-m_f^2)^2}{(p^2-m_f^2)^8}\eB^5,
\eea
\bea
\mathcal{G}_2(p,B)\equiv\frac{\eB^2}{(p^2-m_f^2)^4}-4\frac{3\pt^2-2(\pp^2-m_f^2)}{(p^2-m_f^2)^7}\eB^4-8\frac{45\pt^4+78(\pp^2-m_f^2)+17(\pp^2-m_f^2)^2}{(p^2-m_f^2)^{10}}\eB^6,
\eea
\label{Gdefinitions}
\end{subequations}
so that a rearrangement of terms gives the result shown in Eq.~(\ref{iSfinalfromLandau}).
\end{widetext}


\end{document}